\documentclass[12pt,preprint]{aastex}

\newcommand{\sect}[1]{\S\,\ref{#1}}

\newcommand{\msun}{\ensuremath{\, {M}_\odot}}
\newcommand{\be}{\begin{displaymath}}
\newcommand{\ee}{\end{displaymath}}
\newcommand{\bea}{\begin{eqnarray}}
\newcommand{\eea}{\end{eqnarray}}
\newcommand\msol{{M_{\odot}}}
\newcommand\mstar{{M}}
\newcommand\mr{{M}_r}

\newcommand\cc{$^{12}$C/$^{13}$C}
\newcommand\eli{\log\varepsilon(^7\mbox{Li})}

\newcommand\dlt{\Delta\log\,T}
\newcommand\dm{D_{\rm mix}}
\newcommand\cs{\,cm$^2$\,s$^{-1}$}
\newcommand\kms{\,km\,s$^{-1}$}
 
\shortauthors{Denissenkov et al.}
\shorttitle{Enhanced Extra Mixing: Tidally Locked Binaries}

\begin{document}

\title{FROM CANONICAL TO ENHANCED EXTRA MIXING IN LOW-MASS RED GIANTS:
TIDALLY LOCKED BINARIES}

\author{Pavel A. Denissenkov\altaffilmark{1,2,3}, Brian Chaboyer\altaffilmark{1}, \& Ke Li\altaffilmark{1}}
\altaffiltext{1}{Department of Physics \& Astronomy, Dartmouth College,
   6127 Wilder Laboratory, Hanover, NH 03755-3528, USA;
   Brian.Chaboyer@Dartmouth.edu, Ke.Li@Dartmouth.edu.}
\altaffiltext{2}{Present address: Department of Astronomy, The Ohio State University, 4055 McPherson Laboratory,
       140 West 18th Avenue, Columbus, OH 43210; dpa@astronomy.ohio-state.edu.}
\altaffiltext{3}{On leave from Sobolev Astronomical Institute of St. Petersburg State University,
   Universitetsky Pr. 28, Petrodvorets, 198504 St. Petersburg, Russia.}
 
\begin{abstract}

Stellar models which incorporate simple diffusion or
shear induced mixing are used to describe canonical extra mixing in
low mass red giants of low and solar metallicity.  These models
are able to simultaneously explain the observed Li and CN abundance
changes along upper red giant branch (RGB) in field low-metallicity
stars and match photometry, rotation and \cc\ ratios for
stars in the old open cluster M67.  The shear mixing model
requires that main sequence (MS) progenitors of upper RGB
stars possessed rapidly rotating radiative cores and that specific
angular momentum was conserved in each of their mass shells during
their evolution.  We surmise that solar-type stars will not
experience canonical extra mixing on the RGB because their more
efficient MS spin-down resulted in solid-body rotation, as revealed by
helioseismological data for the Sun.  Thus, RGB stars in 
the old, high metallicity cluster NGC\,6791 should show no evidence for
mixing in their \cc\ ratios.

We develop the idea that canonical extra mixing in a giant component
of a binary system may be switched to its enhanced mode with much
faster and somewhat deeper mixing as a result of the giant's tidal
spin-up.  This scenario can explain photometric and
composition peculiarities of RS CVn binaries.  The tidally enforced
enhanced extra mixing might contribute to the star-to-star
abundance variations of O, Na and Al in globular clusters. This idea 
may be tested with observations of \cc\ ratios and CN abundances in RS
CVn binaries.

\end{abstract} 

 
\section{Introduction}
\label{sec:intro}

\begin{figure}
\plotone{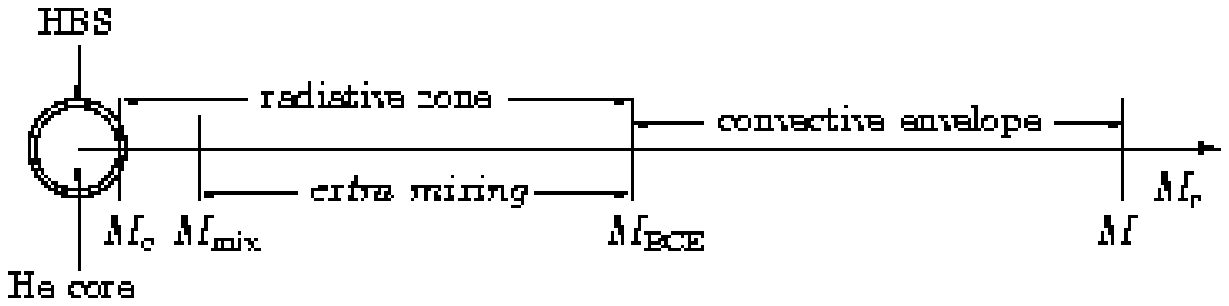}
\caption{Schematic structure of a red giant with extra mixing in its radiative zone.
         HBS and BCE stand for the ``hydrogen burning shell'' and
         the ``bottom of the convective envelope''.
        }
\label{fig:f1}
\end{figure}

Standard stellar evolution models assume that mixing only occurs in
convective regions of stars.  However, there is convincing
evidence that a majority of low-mass ($0.8\la M/\msol\la 2$) stars are
experiencing nonconvective {\it extra mixing} on the upper red giant branch
(RGB) (e.g.,
\citealt{sm79,lea86,sea86,vs88,st92,ch94,ch95,chea98,chdn98,grea00,bea01,gea02,dv03a,sm03,sh03}).
These red giants have a helium electron-degenerate core, a hydrogen
burning shell (HBS) atop the core, and a thin radiative zone between
the HBS and the bottom of the convective envelope (BCE)
(Fig.~\ref{fig:f1}).  The most convincing observational evidence of
mixing in red giants is the evolution of their surface carbon isotopic
ratio \cc\ and abundances $\eli$, [C/Fe] and [N/Fe]\footnote{We use
the standard spectroscopic notations: [A/B]\,=\,$\log\,(N({\rm
A})/N({\rm B}))-\log\,(N({\rm A})/N({\rm B}))_\odot$, where $N({\rm
A})$ and $N({\rm B})$ are number densities of the nuclides A and B;
$\eli = \log (N(^7\mbox{Li})/N(\mbox{H})) + 12$.}  with increasing
luminosity $L$ along the RGB ({\it circles} in Fig.~\ref{fig:f2}). The
first abundance changes that occur on the subgiant and lower red giant
branch ($0.6\la\log L/L_\odot\la 1.0$) are produced by the deepening
BCE during the standard first dredge-up episode (\citealt{i67}). Here, the convective
envelope engulfs the stellar layers in which the CN-cycle had altered
the relative abundances of $^{12}$C, $^{13}$C and N while on the main
sequence (MS). Convection also dilutes the envelope abundance of Li
that has survived only in a thin subsurface layer where the
temperature was less than $\sim$\,$2.5\times 10^6$\,K. At the end of
the first dredge-up, the BCE stops to deepen and begins to retreat.
It leaves a discontinuity of the chemical composition at the level of
its deepest penetration.  The evolutionary changes of the surface
element abundances resume precisely at the moment when the HBS,
advancing in mass inside red giants, erases the composition
discontinuity left behind by the BCE. The commonly accepted
interpretation of this is based on the following assumptions: (1) some
extra mixing is present in the red giants' radiative zones; (2) it
connects the BCE with the layers adjacent to the HBS
(Fig.~\ref{fig:f1}), where the CN-cycle is at work and Li is strongly
depleted, thus dredging up stellar material enriched in N and
deficient in C and Li; (3) at lower luminosities, extra mixing is
shielded from reaching the vicinity of the HBS by a large gradient of
the mean molecular weight (a $\mu$-gradient barrier) built up by the
composition discontinuity (\citealt{chea98,dv03a}).

When the HBS meets and erases the $\mu$-gradient barrier, the
evolution of red giants slows down for a while. This results in
prominent bumps in the differential luminosity functions of globular
clusters, which have really been observed (e.g.,
\citealt{zea99,mea02,cea02,sea02,rea03}).  Therefore, the luminosity
at which extra mixing starts to manifest itself (at $\log
L/L_\odot\approx 1.8$\,--\,2.0 in Fig.~\ref{fig:f2}) is called ``the
bump luminosity''. It divides the RGB into its lower and upper
part. In stars with $M\ga 2\,\msol$, the HBS fails to cross the
$\mu$-gradient barrier before the core helium ignition is
triggered. In concordance with this, no manifestations of extra mixing
have been reported in intermediate-mass red giants (\citealt{g89}).

In the absence of a reliable theory of extra mixing in upper RGB
stars, semi-empirical models are worth applying. In particular, a
simple diffusion model has shown that in the majority of upper RGB
stars the depth and rate of extra mixing do not seem to vary greatly
from star to star.  According to \cite{dv03a}, these quantities can be
parameterized by any pair of correlated values within the close limits
specified by $\dlt\approx 0.19$ and $\dm\approx 4\times 10^8$\cs, to
$\dlt\approx 0.22$ and $\dm\approx 8\times 10^8$\cs.  Here, $\dlt$ is
a difference between the logarithms of temperature at the base of the
HBS and at a specified maximum depth of extra mixing $\mstar_{\rm mix}$
(Fig.~\ref{fig:f1}), and $\dm$ is a constant diffusion
coefficient. 

Given that the majority of metal-poor upper RGB stars,
both in the field and in globular clusters, experience extra mixing
with almost the same depth and rate, \cite{dv03a} have proposed
to call this universal process ``canonical extra mixing''.
Although its physical mechanism is not identified yet, the phenomenon of
Li-rich K-giants and its interpretation are likely to support a rotational mechanism.
Indeed, most of the Li-rich giants are located above the bump luminosity
(\citealt{chb00}). Their proportion among rapid rotators ($v\sin\,i\geq 8$\,\kms)
is $\sim$\,50\%. Compare this to $\sim$\,2\% of Li-rich stars among the much more common slowly
rotating ($v\sin\,i\la 1$\,\kms) K-giants (\citealt{dea02}).
Because for extra mixing driven by kinetic energy of rotation $\dm\propto v^2$,
a tenfold increase of $v$ would enhance $\dm$ from its canonical 
value of $\sim 10^9$\,\cs\ to $\sim 10^{11}$\,\cs. \cite{dh04}
have shown that precisely such enhancement of $\dm$ is required to
produce Li in K-giants via the $^7$Be-mechanism (\citealt{cf71}).
Therefore, following \cite{dv03a}, we hypothesize that sometimes,
when an upper RGB star gets spun up by an external source of angular momentum,
the presumably rotational extra mixing in its radiative zone
may switch from its canonical mode to {\it an enhanced mode} with much faster
and somewhat deeper mixing. 

\cite{dw00} have proposed that
single Li-rich giants might have been spun up by engulfed massive
planets.  Alternatively, a red giant can be spun up by a close stellar
companion.  In this case, its rotation is accelerated by a drag force
originating from viscous dissipation of kinetic energy of a tidal hump
raised in its convective envelope by a companion star. As a result of
this interaction, spin and orbital rotation of the red giant get
synchronized, and its initially elliptical orbit can become
circular. Fig.~\ref{fig:f3} illustrates tidal synchronization in real
stars. Here, we have plotted observational data of \cite{dmea02} on
projected rotational velocities $v\sin i$ and orbital periods $P$ of G
and K giant components of field binaries with available orbital
parameters.  The values of $v\sin i$ were multiplied by the factor
$4/\pi\approx 1.27$ that takes into account a random
orientation of the stars' axes of rotation.  {\it Filled circles} are
binary systems with a circular or nearly circular orbit (those with
eccentricities $e\leq 0.10$). Systems with eccentric orbits ($e >
0.10$) are represented by {\it open circles}.  A set of theoretical
curves is constructed by us under the assumption that $\Omega =
\omega$, where $\Omega$ and $\omega$ are the spin and orbital angular
velocity of a red giant, for 4 values of the red giant's radius:
$R/R_\odot =$\,5, 10, 20 and 40 ({\it solid curves} from left to
right). The tidal circularization time is longer than the
synchronization time (e.g., \citealt{hjea02}), therefore the giants
represented by filled circles are most likely to have synchronized
their rotation.

The goal of this paper is twofold.
First, in \sect{sec:single} we demonstrate that canonical extra mixing 
in solar-metallicity red giants in the open cluster M67 can be modeled
using essentially the same parameters as those adjusted for metal-poor stars.
After that, we attempt to model
enhanced extra mixing in  red giant components of
tidally locked (corotating) binaries that are spun up as a result
of tidal synchronization between their spin and orbital rotation
(\sect{sec:binary}). In order to get a self-consistent picture of
switching from canonical to enhanced extra mixing, in
\sect{sec:binary} we make assumptions and use $\Omega$--profiles and
parameters of extra mixing compatible with those discussed in
\sect{sec:single}. Finally, \sect{sec:concl} contains a summary of our
main conclusions along with a proposal of observational tests
that could confirm or reject our models.

\section{Stellar Evolution Code with Rotation and Extra Mixing}
\label{sec:stevrotmix}

Evolution of rotating stellar models with extra mixing on the upper
RGB is calculated from the zero-age MS (ZAMS) to the RGB tip with an
upgraded version of the computer code used by Denissenkov \& VandenBerg (2003a,b).  
The most recent update is the adoption of Alan Irwin's improved EOS\footnote{We
use the EOS code that is made publicly available at {\tt
ftp://astroftp.phys.uvic.ca/pub/
irwin/eos/code/eos\_demo\_fortran.tar.gz} under the GNU General Public
License.}. In addition, the energy losses due to neutrino emission are
now calculated with the code distributed by \citet{iea96}.  Our
stellar evolution code uses OPAL opacities (\citealt{ri92}) for
temperatures above $\sim$\,$10^4$ K, complemented by \citet{af94} data
for lower temperatures: both datasets assume solar abundances of
$\alpha$--elements.  With the new EOS, we use a value of $l_{\rm
conv}/H_P = 1.75$ for the ratio of the convective mixing length to the
pressure scale height derived from a calibrated solar model.

In our MS models with $M > 1\,\msol$, some overshooting has been
introduced beyond the formal convective core border determined by the
Schwarzschild criterion by assuming that the ratio of the radiative
and adiabatic temperature gradients (logarithmic and with respect to
pressure) at the core border $\nabla_{\rm rad}/\nabla_{\rm ad} = 0.9$
instead of 1.  Using this approach, we succeeded in fitting our
solar-metallicity $4\times 10^9$ yr isochrone\footnote{Necessary
bolometric corrections and $T_{\rm eff}$\,--\,color transformations
were calculated with a code kindly presented to one of us (PAD) by Don
VandenBerg.}  (Fig.~\ref{fig:f4}, {\it solid curve}) for the high
precision color-magnitude diagram (CMD) of M67 published by \cite{s04}
(Fig.~\ref{fig:f4}, {\it open circles}).  We ignored any microscopic
mixing, such as atomic diffusion, gravitational settling and radiative
levitation. 

The treatment of extra mixing in the radiative zones separating the BCE
from the HBS is the same as that described by \citet[][see their
equation (1)] {dv03a}.
Effects of shellular rotation on the internal structure and evolution of stars
are treated in the same way as by \cite{dv03b}. 
The code employed in this work only
considers the simplest cases of the angular momentum profile evolution: either
(a) supporting the uniform (solid-body) rotation, $\Omega(t,\mr)=\Omega(t)$, 
or (b) keeping the constant specific angular momentum
in each mass shell, including convective regions, $j(t,\mr)\equiv\frac{2}{3}r^2\Omega =j(0,\mr)$. 
These two opposite cases correspond to
a maximum possible and a zero efficiency of the angular momentum redistribution
throughout a star. 

\section{Canonical Extra Mixing in Single Red Giants of Solar Metallicity}
\label{sec:single}

In this section, we present results of our new stellar evolution computations
that, when being compared with recent observational data on photometry, rotation
and chemical composition of Population I (in the open cluster M67) and Population II (both in the field
and in globular clusters) low-mass stars, allow to constrain the strength of
rotation-induced shear mixing.

\subsection{Basic Equations and Assumptions in the Shear Mixing Model}

When studying the evolution of metal-poor stars with $\mstar\approx 0.9\,\msol$, \cite{dt00} and \cite{dv03a} have
proposed that canonical extra mixing in upper RGB stars can be identified with 
turbulent diffusion induced by secular shear instability in their differentially
rotating radiative zones and that it
can be modeled using the coefficient of vertical turbulent diffusion derived by \cite{mm96}
\bea
D_{{\rm v}} = f_{\rm v}\times
8Ri_{\rm c}\frac{K}{N_T^2}\left[\frac{1}{5}\Omega^2\left(\frac{\partial\ln\Omega}{\partial\ln r}\right)^2-N_\mu^2\right].
\label{eq:dv}
\eea
Here,
\be
N_T^2 = \frac{g\delta}{H_P}(\nabla_{{\rm ad}}-\nabla_{{\rm rad}}),
\ \ \mbox{and}\ \
N_\mu^2 = g\varphi\,\left|\frac{\partial\ln\mu}{\partial r}\right|
\ee
are the $T$- and $\mu$-components of the square of the
Brunt--V\"{a}is\"{a}l\"{a} (buoyancy) frequency,
$g$ is the local gravity,
$Ri_{\rm c} = \frac{1}{4}$ is the classical critical Richardson number, and
\bea
K = \frac{4acT^3}{3\kappa\rho^2C_P}
\eea
is the thermal diffusivity with $\kappa$ and $C_P$ representing the opacity and
the specific heat at constant pressure, respectively. The quantities
$\delta =
-\left(\partial\ln\rho/\partial\ln T\right)_{P,\mu}$ and
$\varphi = \left(\partial\ln\rho/\partial\ln\mu\right)_{P,T}$
are determined by the equation of state.
                                                                                                                                                          
Equation (\ref{eq:dv}) gives simultaneously the extra mixing rate
\bea
D_{{\rm mix}} \approx
f_{\rm v}\times\frac{8}{5}Ri_{\rm c}\frac{K}{N_T^2}\Omega^2\left(\frac{\partial\ln\Omega}{\partial\ln r}\right)^2
\label{eq:dvmix}
\eea
and depth $\mstar_{\rm mix}$, the latter being determined as the coordinate of
the first mass shell beneath the BCE where $D_{\rm v} = 0$, or
\bea
N_\mu^2 \equiv g\varphi\,\left|\frac{\partial\ln\mu}{\partial r}\right| =
\frac{1}{5}\Omega^2\left(\frac{\partial\ln\Omega}{\partial\ln r}\right)^2.
\label{eq:dvmmix}
\eea
Although the depth given by equation (\ref{eq:dvmmix}) for reasonable $\Omega$--profiles
is surprisingly close to that specified by the parameter $\dlt = 0.19$\,--\,0.22 in our simple diffusion model,
the rate calculated with equation (\ref{eq:dvmix}) when $f_{\rm v}=1$
is found to be too slow for canonical extra mixing (\citealt{dv03a}).
Therefore, we have to insert {\it an enhancement factor} $f_{\rm v}\gg 1$ into equations (\ref{eq:dv}) and (\ref{eq:dvmix}).

For the low values of $v\sin i$ usually measured in MS stars with $M\la 1\,\msol$,
rotational mixing, such as meridional circulation or turbulent diffusion, can reproduce
the evolutionary changes of [C/Fe] in their upper RGB descendants only assuming
that the specific angular momentum is conserved in each mass shell of these stars
during their entire evolution from the ZAMS to the RGB tip
(\citealt{sm79,st92,dt00,dv03a}). Even having assumed that,
some differential rotation in the initial models is still required.
Following Denissenkov \& VandenBerg (2003a,b),
we set up such differential rotation in our ZAMS models
assuming that the ratio of the equatorial centrifugal acceleration to the gravity is
constant along the radius and that it is a small fraction of the critical ratio
for the Roche potential used in our rotating stellar models; i.e.,
\bea
\varepsilon(\mstar_r)\equiv \frac{r^3\Omega^2}{3GM_r}=f_\varepsilon\,\varepsilon_{\rm crit}=\mbox{constant}, 
\label{eq:initrot}
\eea
where $\varepsilon_{\rm crit}\approx 0.24$.
The ratio of the angular velocity near the center to $\Omega$ at the surface obtained
in this way is less than 10 for all of the ZAMS models considered below.
Thus, we admit that the solid-body rotation
of the inner Sun as revealed by helioseismological data (e.g., \citealt{chapea99}) may not necessarily be
applicable to all low-mass MS stars.
The free parameter $f_\varepsilon$ is chosen so as to
get theoretical  values of $v_{\rm rot}$ close to projected rotational velocities of
MS and subgiant stars.  After that we evolve our
models from the ZAMS to the bump luminosities keeping $j$ constant in every mass shell.
No extra mixing is allowed in the models until they will reach the bump luminosities.
For the sake of simplicity, we ignore mass loss (in fact, it becomes important only in the very vicinity of the RGB tip).

\subsection{A Summary of Results for the Single Star Case}
                                                                                                                                                          
Equations (\ref{eq:dvmmix}) and (\ref{eq:dvmix})
with values of $\Omega$ and $(\partial\ln\Omega/\partial\ln r)$ taken from the radiative zones of our bump luminosity
models give approximately the same depths of extra mixing
and diffusion coefficients that are only $\sim$\,15\,--\,25 times as small
as those constrained by our semi-empirical diffusion model.
This conclusion is true not only for metal-poor low-mass stars, as was anticipated
by \cite{dv03a}, but also for their solar-metallicity counterparts.
To demonstrate this, we have computed the evolution of three
rotating stellar models: the first two for the same mass $M=0.85\,\msol$ and helium abundance $Y=0.24$ but
for two different heavy-element mass fractions $Z=0.002$ and $Z=0.0005$, and the third model for
$M=1.35\,\msol$, hydrogen abundance $X=0.70$ and the solar value of $Z=0.0188$.
                                                                                                                                                          
The initial rotation profiles of the first two models have been specified by the parameter $f_\varepsilon = 0.0003$.
This results in surface equatorial rotational velocities $v_{\rm rot}$
monotonously decreasing from $\sim$\,7\kms\ to $\sim$\,4\kms\
during the models' MS lives. When these metal-poor models leave the MS and become subgiants,
their rotation slows down to $v_{\rm rot} \leq 4$\kms\ due to a surface radius increase
(Fig.~\ref{fig:f5}, {\it solid curve}).
Similar values of ``true'' mean rotational velocities have been measured
in the MS turnoff and subgiant stars in the globular clusters
NGC 104, NGC 6397 and NGC 6752 by \cite{lg03} ({\it squares with errorbars} in Fig.~\ref{fig:f5}).
The third model evolves along
the RGB of the $4\times 10^9$ yr isochrone that matches the CMD of
the old solar-metallicity open cluster M67 (Fig.~\ref{fig:f4}, {\it dashed curve}).
For this model, the parameter $f_\varepsilon = 0.00075$ has been chosen so as the computed behavior of $v_{\rm rot}$
along the rotational isochrone of the same age constructed using
stellar models with $M \geq 1.1\,\msol$ and the same parameter $f_\varepsilon$
(Fig.~\ref{fig:f6}, {\it solid curve})
to be in a qualitative agreement with the observed decrease of $(4/\pi)v\sin\,i$ in the M67 stars
whose evolutionary status changes from the MS to the subgiant branch (Fig.~\ref{fig:f6}, {\it circles}).
Unfortunately, at present we cannot discriminate between the cases of differential and solid-body rotation
of M67 stars (compare {\it solid} and {\it short-dashed curves} in Fig.~\ref{fig:f6}).
However, the observational data in Fig.~\ref{fig:f6} seem to support
our assumption of slow surface rotation ($v_{\rm rot}\approx 10$\,\kms) of M67 stars on the early MS
(compare {\it solid} and {\it short-dashed curves} with {\it dot-long-dashed} and {\it long-dashed curves}).
The stars from Fig.~\ref{fig:f6} are also plotted in Fig.~\ref{fig:f4} as {\it filled circles}.
Comparison of these two figures shows that the stars with $(B-V)_0 \approx 0.55$\,--\,0.60 in Fig.~\ref{fig:f6}
have $\mstar < 1.35\,\msol$. Their masses are probably close to the M67 MS turn-off mass,
which is $\mstar\approx 1.25\,\msol$. Only the objects with $(B-V)_0\ga 0.9$ in Fig.~\ref{fig:f6}
are subgiants and lower RGB stars having $M\approx 1.35\,\msol$.

In Fig.~\ref{fig:f7}, {\it vertical dotted and dashed line segments} in panels d
indicate depths specified by $\dlt = 0.19$ and 0.22 in the
third ($\mstar = 1.35\,\msol$, $Z=0.0188$) unmixed model that is located immediately above
its bump luminosity.  {\it Solid line segment} points to the mixing depth calculated
with equation (\ref{eq:dvmmix}) using an $\Omega$--profile extracted from the same stellar model.
Note the proximity of the broken and solid line segments.
                                                                                                                                                          
Starting from the bump luminosity models, we continue our stellar
evolution computations.  Rotational shear mixing described by equation
(\ref{eq:dv}) is now taken into account.  To be more exact, we
introduce a diffusion coefficient $D_{\rm mix} = D_{\rm v}$ for $D_{\rm v} > 0$
while letting $D_{\rm mix} = 0$ for $D_{\rm v}\leq 0$.  A new
$\Omega$--profile required to recalculate $D_{\rm v}$ is taken from the
evolving rotating stellar models in each time step. As in the models
below the bump luminosities, the specific angular momentum is still
assumed to remain constant in each mass shell, including convective
envelopes.  Thus, our new extra mixing computations include for the
first time both a plausible physical description for the diffusion
coefficient $D_{\rm mix}$, with its profile allowed to change along the
whole length of upper RGB\footnote{\cite{dv03a} used a fixed $\dm$-profile
proportional to $D_{\rm v}$ in the bump luminosity model.}, 
and a feedback effect of rotation and chemical mixing on
the structure and evolution of red giants.  In order to simulate the
evolutionary variations of the surface \cc\ ratio and abundances of
Li, C and N in the field metal-poor upper RGB stars we had to increase
by a factor of $f_{\rm v}=15$\,--\,25 the value of $D_{\rm v}$ given
by \cite{mm96} in our $0.85\,\msol$ rotating models
(Fig.~\ref{fig:f2}). At the same time, the \cc\ ratios in a couple of
upper RGB stars and in a few clump stars in the cluster M67 were
reproduced by our $1.35\,\msol$ model only after $D_{\rm v}$ had
been multiplied by $f_{\rm v}=15$ (Fig.~\ref{fig:f8}, {\it dashed
curve}).  For comparison, {\it dotted curve} in Fig.~\ref{fig:f8} has
been computed with the diffusion model using $\dlt = 0.22$ and $\dm =
8\times 10^8$\cs.  Also, compare our Fig.~\ref{fig:f2} with Fig.~4
from \cite{dv03a}. The ratio between the enhancement factors 25 and
15, adjusted for the low-mass red giant models of different
metallicities, is even smaller than the one between our empirically
constrained limits for the diffusion coefficient: $\dm = 8\times
10^8$\,\cs\ and $\dm = 4\times 10^8$\,\cs.
                                                                                                                                                          
Fig.~\ref{fig:f7} shows element abundance profiles, $\mu$-gradients and diffusivities
for two rotating stellar models with $\mstar = 1.35\,\msol$,
$Z=0.0188$ and $f_\varepsilon = 0.00075$ located close to their bump luminosity.
Compared to the case of metal-poor red giants, the only important qualitative distinction here
is the additional hump in the $\mu$-gradient
profile at $\mstar_r/\msol \approx 0.2535$ in the upper RGB model clearly seen in panel d. It is built up by H abundance
variations that accompany the $^{12}$C to $^{14}$N transformation (see panel b).
Because of this hump, the vertical turbulent diffusion cannot approach as close to the HBS as in
the low-metallicity case.
It can hardly reach a peak in the $^{13}$C abundance
profile (panel b). At the same location, the $^{12}$C depletion just sets in.
This may explain why in the solar-metallicity case canonical extra mixing only produces modest changes of
the surface \cc\ ratio (Fig.~\ref{fig:f8}).
                                                                                                                                                          
To summarize, multiplying the diffusion coefficient $D_{\rm v}$ from \cite{mm96} by a factor of
$f_{\rm v}=20\pm 5$, we can satisfactorily
reproduce the observed evolutionary variations of the surface \cc\ isotopic ratio and abundances of
Li, C and N in the field metal-poor upper RGB stars (Fig.~\ref{fig:f2}) as well as
the scanty \cc\ data for the upper RGB and clump stars in the old solar-metallicity open cluster M67
(Fig.~\ref{fig:f8}). This can only be done under the assumption of
differential rotation both in the low-mass MS ancestors of these evolved stars and in the convective
envelopes of red giants.
Otherwise, one has to search for an alternative mechanism of canonical extra mixing.

Interestingly, our two models of canonical extra mixing (the diffusion model and the shear instability model)
predict qualitatively different evolutionary changes of the surface Li abundance near the RGB tip.
Whereas in the first model $\eli$ remains very low, 
in the second model $\eli$ resumes its post-first-dredge-up value
and even exceeds it towards the RGB tip (compare upper panels in our Fig.~\ref{fig:f2} and in Fig.~4 from \cite{dv03a}).
This difference is due to a considerable increase of the diffusion coefficient $D_{\rm v}\propto K$ (equation (\ref{eq:dv}))
in the second model caused by a growth of the thermal diffusivity $K$ with the luminosity and the assumed constancy of
$\dm$ in the first model. The larger diffusion coefficient leads to some Li production via the $^7$Be-mechanism.
This new result can explain the higher Li abundances in clump stars compared to those in red giants discussed by
\cite{pea01}.

Now, we want to comment on the required enhancement of $D_{\rm v}$.
We can formally comply with this requirement by choosing a value of the critical Richardson
number $Ri_{\rm c}$ larger than $\frac{1}{4}$ instead of taking $f_{\rm v}\gg 1$. 
Indeed, the modern models of shear-driven turbulence do predict
that $Ri_{\rm c}$ should be at least 4 (\citealt{c02}) or even 6.4 (\citealt{bh01}) times as large as its classical
value. If we accept this then the enhancement factor $f_{\rm v}$ in equation (\ref{eq:dv})
has only to be as large as $\sim 20/6.4\approx 3$ for our model to reproduce the observations.
It should also be noted that equation (\ref{eq:dv}) takes into consideration only the largest
turbulent eddies. Allowing for eddies from the whole turbulent spectrum would increase
the coefficient $D_{\rm v}$ in a similar way as the Full Spectrum of Turbulence model (\citealt{cm91})
generates much more vigorous convection than the Mixing Length Theory.

As regarding the depth of shear mixing (\ref{eq:dvmmix}), it
is directly constrained by the kinetic energy of differential
rotation available to do work against the buoyancy force.
It is our assumption of $\Omega$ increasing with depth
in the convective envelopes of low-mass red giants and in their MS 
ancestors that allows shear mixing to approach close enough to the HBS.
In order to reconcile this assumption with the well established fact of
the Sun's solid-body rotation, we refer to the recent work of \cite{tch04}, according to which,
in a solar-type star, angular momentum can be extracted from its radiative core by
internal gravity waves generated in its convective envelope. The efficiency of this
process should depend on the thickness of convective envelope. Therefore,
we hypothesize that rotation-induced extra mixing may only work in those red giants whose MS ancestors had
shallower convective envelopes than the Sun. Indeed, both our ZAMS models, the one with $\mstar = 1.35\,\msol$ and $Z=0.0188$
and that with $\mstar = 0.85\,\msol$ and $Z=0.0005$
possess much thinner convective envelopes than the solar ZAMS model ($\log M_{\rm ce}/\msol = -3.92$,
$-2.32$ and $-1.62$, respectively).
Thus, we conjecture that metal-rich stars with $\mstar\la 1.2\,\msol$ might not experience canonical extra mixing
on the RGB at all because their MS ancestors had thick surface convective zones and, as a result, nearly solid-body internal rotation,
like the Sun. To test this hypothesis, it is necessary to determine \cc\ ratios in upper RGB stars of an open cluster older than M67.
A good candidate for these purposes is the open cluster NGC\,6791. It has [Fe/H]\,$=+0.4$ and an age of $(8.0\pm 0.5)\times 10^9$ yr
(\citealt{chea99}).

\section{Enhanced Extra Mixing in Giant Components of Tidally Locked Binaries}
\label{sec:binary}

Given that enhanced extra mixing is deeper than
canonical one, it may penetrate the layers in outskirts of the HBS in which not only $^{12}$C
but also $^{16}$O is depleted in the CNO cycle and where $^{23}$Na is produced in the reaction
$^{22}$Ne(p,$\gamma)^{23}$Na (\citealt{dv03a}).
Therefore, enhanced extra mixing in the extinct upper RGB stars that were slightly more massive
($0.9\la \mstar/\msol\la 2$) than the MS turn-off stars in the present-day
globular clusters ($\mstar\approx 0.8$\,--\,$0.9\,\msol$) might have been one of the yet unidentified primordial sources of
the global O--Na anticorrelation in globular cluster stars (\citealt{dw04}).

For the binary star case, we consider two basic stellar models: a solar-metallicity one with $\mstar = 1.7\,\msol$, and a metal-poor one with
$\mstar = 1.0\,\msol$ and $Z=0.0005$. The first model has the mass and metallicity typical for both Li-rich K-giants and a sample of
chromospherically active late-type giants --- primaries of the RS Canum Venaticorum (CVn) binaries 
whose relevance to the problem of enhanced extra mixing will be discussed later.
The second model represents a low-mass star that had completed its life in the past
in a moderately metal-poor globular cluster. Both models will be ``placed'' into close binary systems with less massive stellar companions.
Our main goal is to find out to what noticeable photometric and composition anomalies the tidal spin-up of these models may lead on the RGB.

\subsection{Description of Tidal Interaction in a Binary System}

\subsubsection{Basic Equations}

If the initial separation between low-mass stellar components of a binary system
is too small then the primary star can fill its Roche lobe before
having reached the RGB tip. In this case, the radius of the red giant primary has an upper limit constrained by the Roche lobe radius
$R\leq R_L = E(q) a$, where $a$ is the binary semi-major axis, $q=m/M$ is its mass ratio (secondary over primary), and, according
to \cite{e83},
\be
E(q)\approx \frac{0.49}{0.6+q^{2/3}\ln (1+q^{-1/3})}.
\ee
{\it Dashed curve} in Fig.~\ref{fig:f3} delineates the maximum surface rotational velocity of a red giant that fills its Roche
lobe in a binary system with $\mstar = 1.7\,\msol$, $q=0.5$, and
an orbital period $P$, the latter being related to $a$ via Kepler's third law
\bea
\left(\frac{a}{R_\odot}\right) = 4.207 P_{\rm d}^{2/3}\left(\frac{\mstar}{\msol}\right)^{1/3}(1+q)^{1/3},
\label{eq:k3}
\eea
where $P_{\rm d}$ is the period in days.
We see that all stars but one lie below this curve. The only outlier
is the G5\,III star HD\,21018. It has $v\sin i = 22.7$\,\kms, and the lithium abundance $\eli = 2.93$ (\citealt{h78}).
It is most likely to be an intermediate-mass ($\mstar\ga 2.5\,\msol$) star
crossing the Hertzsprung gap on its way from the MS to the RGB region.
Its location on the CMD ({\it asterisk} in Fig.~\ref{fig:f9}) confirms our
hypothesis. Thus, it may still preserve its initial
high Li abundance, while the large value of $v\sin i$ may be a signature of its previous fast rotation characteristic of the MS B-type stars.

In our work, the tidal evolution of orbital and rotational parameters of binary star systems is modeled using
differential equations derived by \cite{h81}. Hut has elaborated upon the weak friction model of \cite{a73},
which assumes that the tidal humps in a gravitationally distorted rotating star have their equilibrium shape,
as described by \cite{z77}, but with a constant time lag between the moment when the tidally deformed equipotential
surface has been formed in the absence of dissipation and the current binary star configuration.
For the sake of simplicity, we only consider the tidal evolution of primary stars possessing
convective envelopes in binary systems with the zero eccentricity.
In this case, only the spin rotation of the primary and the binary semi-major axis are affected by the tidal interaction,
their changes being controled by the following equations:
\bea
\frac{d\Omega}{dt} & = & 3\frac{k}{t_{\rm F}}\frac{q^2}{r_{\rm g}^2}\left(\frac{R}{a}\right)^6(\omega - \Omega),
\label{eq:domdt} \\
\frac{da}{dt} & = & -6\frac{k}{t_{\rm F}}q(1+q)\left(\frac{R}{a}\right)^8a\,\frac{\omega - \Omega}{\omega},
\label{eq:dadt}
\eea
where
\bea
t_{\rm F}\approx t_{\rm conv}\approx \left(\frac{\mstar R^2}{L}\right)^{1/3} =
0.4304\left[\frac{(\mstar/\msol)(R/R_\odot)^2}{(L/L_\odot)}\right]^{1/3}\ \mbox{yr}
\label{eq:tf}
\eea
is a characteristic time of viscous friction, which approximately equals the convective time scale
(\citealt{z77}) (friction experienced by the tidal humps is produced by
the convective turbulent viscosity in the envelope),
$L$ is the primary star's luminosity, $r_{\rm g}$ is its radius of gyration that determines
the total angular momentum  $I=r_{\rm g}^2MR^2$, 
$\Omega$ and $\omega$ are its spin and orbital angular velocities.

Equations (\ref{eq:domdt}\,--\,\ref{eq:dadt}) contain the apsidal motion constant of the primary
\bea
k = \frac{3-\eta[a(R)]}{2\{2+\eta[a(R)]\}}.
\label{eq:k}
\eea
The structural function $\eta(a)$ can be obtained by integrating the differential equation of Radau
(e.g. \citealt{cw02}):
\bea
a\,\frac{d\eta}{da} + 6\,\frac{\rho}{\overline{\rho}}\,(\eta + 1) + \eta(\eta - 1) = 6
\label{eq:radau}
\eea
with the boundary condition $\eta(0)=0$.
Here, $\rho(a)$ is the primary's density profile, and $\overline{\rho}(a)$ is the average density inside the equipotential
surface (isobar) whose rotationally distorted shape\footnote{According to \cite{kt70}, within a high accuracy the structural changes of
stars in tidally locked binaries are caused by their rotation and not by tidal forces.
Therefore, we do not take into account the tidal effects on the primary's structure and evolution.}
is described as
\be
r = a[1 - \varepsilon(a)P_2(\cos\theta)]\approx a,
\ee
where $P_2$ is the Legendre polynomial of order 2, $\theta$ is the colatitude, and
$$  
\varepsilon = \frac{\Omega(a)^2a^3}{3GM_P}
$$  
is the ratio of the equatorial centrifugal acceleration to gravity divided by 3, with $M_P$ being the mass
inside the isobar $P$.
The quantities $\eta$ and $\varepsilon$ are related to each other via
$\eta = (d\ln\varepsilon/d\ln a)$.

The quantity $\varepsilon$ has already been used to set up differential rotation in our ZAMS
models (equation (\ref{eq:initrot})). In our solar metallicity ZAMS model with $\mstar = 1.7\,\msol$,
the initial $\Omega$--profile is specified by the same parameter $f_\varepsilon = 0.00075$
as in our single star model with $\mstar = 1.35\,\msol$ in \sect{sec:single}.
It results in the surface rotational
velocity $v_{\rm rot}\approx 10$\,\kms. In fact, MS stars with $\mstar\ga 1.4\,\msol$ have extremely thin
surface convective zones, therefore their rotation is not thought to experience magnetic braking (or other type of braking
whose efficiency depends on the thickness of convective envelope).
Hence, our $1.7\,\msol$ model should have rather been assigned $v_{\rm rot}\approx 100$\,\kms\ instead of 10\,\kms, and it
should have rotated as the solid body. However, this inconsistency does not
matter when we study the evolution of this model star as a primary in a close binary system because, after having been synchronized
with its orbital rotation on the RGB, the primary will acquire a surface rotational velocity by far exceeding that
it would have if it were a single star. So, the information about its previous surface rotation will be completely lost.
What really matters is that our differentially rotating ZAMS model has approximately the same value of $\Omega$
within its inner region, that will be occupied by the radiative zone on the upper RGB,
that it would have if it rotated as a solid body with $v_{\rm rot}\approx 100$\,\kms.
Indeed, in Fig.~\ref{fig:f10} {\it long-dashed horizontal line} shows the uniform rotation profile
with $v_{\rm rot} \approx 100$\kms, while {\it dot-long-dashed curve} presents the $\Omega$--profile obtained using
equation (\ref{eq:initrot}) with $f_\varepsilon = 0.00075$. Both profiles almost coincide at $\mr < 0.4\,\msol$.
Therefore, under the assumption of the conservation of the specific angular momentum $j$ in each mass shell, including convective regions,
radiative zones of the single RGB models that have started their evolution on the ZAMS with these two different
rotation profiles (the uniform one with $v_{\rm rot}\approx 100$\,\kms, and the differential one
with $v_{\rm rot}\approx 10$\,\kms) will possess similar differential rotation.
{\it Dot-short-dashed curve} in Fig.~\ref{fig:f10} demonstrates the rotation profile in the bump luminosity
model obtained from the initial dot-long-dashed profile assuming $j(t,\mr)=j(0,\mr)$ for $0\leq\mr\leq M$. For comparison,
{\it short-dashed horizontal line} gives the uniform $\Omega$--profile resulted from the initial
long-dashed one assuming the solid-body rotation of the whole star during its entire evolution.
In all the cases, the total angular momentum is conserved. Since the bulk of the red giant's angular momentum
is contained in its convective envelope, the short-dashed line also approximates quite well the envelope rotation
(on the right of {\it vertical dotted line}) in the alternative case, that we did not consider, in which $j$ is only
conserved in radiative regions, while $\Omega$ is maintained constant in convective regions. In that case,
as we mentioned, the radiative zone would rotate much slower than in the case represented by 
the dot-short-dashed curve (compare the values of $\Omega$ on the short-dashed and dot-short-dashed
profiles at the BCE).
                                                                                                                                                          
Before ``placing'' our models into binary systems, we have computed their single stellar evolution.
For the metal-poor model ($\mstar = 0.85\,\msol$, $Z=0.0005$), the initial internal $\Omega$--profile has been specified
by the same parameter $f_\varepsilon = 0.0003$ as in \sect{sec:single}.
During the single stellar evolution computations, we were tabulating the quantities $R$, $k$ and $r_{\rm g}$ as functions of
age and luminosity. On the RGB, the ratio $k/r_{\rm g}^2$ has been found to decrease very slightly,
from about 0.5 to nearly 0.2 for both models.
Hence, from equations (\ref{eq:domdt}) and (\ref{eq:tf}) we can estimate the primary's synchronization time 
(the time after which $\Omega\approx\omega$) as
\bea
t_{\rm syn}\approx \frac{t_{\rm F}}{q^2}\left(\frac{a}{R}\right)^6\approx
0.4304\left[\frac{(\mstar/\msol)(R/R_\odot)^2}{(L/L_\odot)}\right]^{1/3}q^{-2}\left(\frac{a}{R}\right)^6\ \mbox{yr}.
\label{eq:tsyn}
\eea
Furthermore, for values of $q\ga 0.5$, the product in front of the last term in equation (\ref{eq:tsyn})
does not deviate much from unity during the entire evolution of our models from the subgiant branch to the RGB tip. Therefore, for
binaries with the mass ratio $0.5\la q < 1$, one can use the simpler estimate $t_{\rm syn}\approx (a/R)^6$\,yr.
Since the low-mass stars spend a time of the order of $10^8$ yr on the RGB, only primary stars in the binaries with $a\la 20\,R$
are expected to get synchronized on this evolutionary stage. If we want our primary to be tidally spun up already on the lower RGB,
where $R\la 10\,R_\odot$, we should only consider the binaries with $a\la 200\,R_\odot$, or, in other words, only those with
$P_{\rm d}\la 300$ (equation (\ref{eq:k3})).

\subsubsection{Choosing a Rotation Profile in the Convective Envelope}

Unfortunately, we do not know how rotation, convection and tidal friction interact with each other in the red giant's
envelope. Even 3D-hydrodynamical simulations will hardly be able to model such complex situation with confidence in the nearest future.
Therefore, we cannot do anything else but to make some plausible assumptions about the outcome of this interaction and to use as many as
possible relevant observational data to constrain our assumptions. We should also choose rotation profiles and
parameters of extra mixing in our binary red giant models consistent with those employed in \sect{sec:single} for the single star case.

In Fig.~\ref{fig:f11}, we have plotted $\Omega$--profiles in our solar-metallicity
$1.7\,\msol$ lower RGB models that started their evolution (on the ZAMS),
either as a single or a binary star, with the same rotation parameter
$f_\varepsilon = 0.00075$, and so with the same value of $v_{\rm rot}$, but for which different assumptions about
the rotation law in their convective envelopes were made.
All the models except that plotted with {\it short-dashed curve} have reached nearly the same evolutionary point
immediately below the bump luminosity. {\it Dot-dashed curve} gives the $\Omega$--profile
in the single star model. {\it Long-dashed, solid} and {\it short-dashed curves} show internal rotation of red giants in binaries
with the same initial parameters: $q=0.5$, $a=80\,R_\odot$. {\it Dotted curve} has $q=0.5$ and $a=50\,R_\odot$.

The long-dashed curve is obtained under the assumption that,
after the corotation has been achieved ($\Omega = \omega$) as a result of tidal interaction in the close binary,
the whole convective envelope of the red giant rotates as the solid body, i.e. $\Omega(r)=\omega$
for $R_{\rm BCE}\leq r\leq R$. However, in this case the radiative zone
would rotate slower than in the single star case (compare the parts of the dot-dashed and long-dashed curves
just beneath the BCE, at $\log R/R_\odot\la -0.2$).
Given that on the RGB hydrogen rich material is slowly moving from the convective envelope to the helium core through the radiative
zone and the specific angular momentum is assumed to be conserved in this zone,
we anticipate that, in the more evolved models, the $\Omega$--profile developed from that represented by the long-dashed curve will
be parallel to the dot-dashed curve (like the short-dashed curve showing the advanced evolution of
the solid curve) but it will entirely lie below it. In that case,
since we assume that $\dm\propto\Omega^2$ (e.g., equation (\ref{eq:dv})), the rate of extra mixing would be reduced compared to
the canonical case, in spite of the red giant's faster surface rotation. This seems to be at odds with
the reported very high percentage of Li-rich giants among rapidly rotating stars (\citealt{dea02}) and
the necessity to enhance $\dm$ from $\sim 10^9$\,\cs\ to $\sim 10^{11}$\,\cs\ in order to enable the $^7$Be-mechanism 
(\citealt{dh04}).
Therefore, we reject the $\Omega$--profile represented by the long-dashed curve as implausible.

In \sect{sec:single}, we have made the assumption that even convective envelopes of single stars 
rotate differentially, with the specific angular momentum
conserved in each mass shell inside them, during the entire evolution from the ZAMS.
Without that assumption rotational mixing driven by the secular shear instability in the radiative zones of upper RGB stars would be too weak
to be identified with canonical extra mixing. At the same time, we admit that tidal interaction in a close binary system with
a red giant component can enforce solid-body rotation of at least an outer part of its convective envelope.
Making an appeal to observational data, we first notice that some of the chromospherically active late-type giants in the RS CVn binaries
(for instance, those shown as {\it filled circles} in Fig.~\ref{fig:f9}) have almost equal orbital and rotational periods.
This means that their surface layers have already been spun up by the tidal drag force to the orbital angular velocity,
$\Omega(R)=\omega$. Although we know nothing about their internal rotation in the convective envelope below the surface,
it is naturally to think that the same tidal force induces corotation of their subsurface convective layers
as well, at least down to some depth $R_{\rm cor}$, i.e. $\Omega(r)=\omega$ for $R_{\rm cor}\leq r\leq R$, where
$R_{\rm BCE} < R_{\rm cor} < R$. We have seen that the assumption of $R_{\rm cor}=R_{\rm BCE}$ (long-dashed curve in
Fig.~\ref{fig:f11}) appears to be wrong if we want to explain canonical extra mixing and the phenomenon of Li-rich
giants assuming that $\dm\propto\Omega^2$ in both cases. 

The idea that $R_{\rm cor}=R$ while $j$\,=\,constant in the convective envelope is not good either because in that case
rotation of the red giant's radiative zone near the HBS $\Omega(R_{\rm c})$ could easily become
supercritical. Indeed, if in the whole radiative zone $j$\,=\,constant as well, then
\bea
\frac{\varepsilon(r)}{\varepsilon_{\rm crit}} \approx 7.3\times 10^{-4}
\frac{(R/R_\odot)^2(v_{\rm rot}/10\,\mbox{km\,s}^{-1})^2}{(\mr/\msol)(r/R_\odot)}.
\eea
The hydrogen burning shell always stays at $r\approx R_{\rm c}\approx 0.02\,R_\odot$,
and the helium core mass in stars located close to the bump
luminosity is $\mr\approx 0.3\,\msol$, which results in
$\varepsilon(r)/\varepsilon_{\rm crit}\approx 0.12\,(R/R_\odot)^2(v_{\rm rot}/10\,\mbox{km\,s}^{-1})^2$.
Thus, in a binary RGB star that has reached a corotation at the surface with $v_{\rm rot}=10$\,\kms
(Fig.~\ref{fig:f3} shows that this velocity is still less than the maximum possible value), rotation near the HBS
could become critical already for $R\approx 2.9\,R_\odot$, i.e. far below the bump luminosity. This disagrees with the presence of
corotating binary red giants close to and even above the bump luminosity.

Equations (\ref{eq:domdt}\,--\,\ref{eq:dadt}) were derived assuming for simplicity that $\Omega$\,=\,constant throughout the convective envelope.
This may be a good assumption for thin convective envelopes but an RGB star has a very thick convective envelope with
$R/R_{\rm BCE}\gg 1$. We can divide it into a number of thin spherical layers and consider them as rotating independently of one another.
This approach is consistent with the assumption of differentially rotating convective envelopes made for single giants.
If the tidal drag force had not varied along the radius inside the convective envelope then it would have naturally been to think that
all the layers get synchronized at the same time. However, the transverse component of the tidal force strongly depends on $r$
(\citealt{h81}),
\bea
F_\theta = 3\,MR\,\frac{k}{t_{\rm F}}\,q^2\left(\frac{r}{a}\right)^7(\Omega - \omega).
\eea
Therefore, the deeper convective layers are most likely to get synchronized much later than the layers near the surface.

In our work, $\Omega$--profiles in convective envelopes of RGB model stars in close binary systems are computed as follows:
{\it (i)} using the values of $R$, $k$ and $r_{\rm g}$ tabulated as functions of 
age and luminosity during the computations of  the single stellar evolution,
we first solve equations (\ref{eq:domdt}\,--\,\ref{eq:dadt}) for a specified set of parameters $q$ and $a$;
this gives us dependences of $v_{\rm rot}$ on $L$ (or $M_V$), and $\Omega/\omega$ on $R$,
like those plotted in Figs.~\ref{fig:f12} and \ref{fig:f13} for our solar metallicity $1.7\,\msol$ model in a binary system
with the mass ratio $q=0.5$ ({\it dashed curves} are obtained for $a=50\,R_\odot$ ($P\approx 25$ days) while
{\it solid curves} are for $a=80\,R_\odot$ ($P\approx 50$ days); in Fig.~\ref{fig:f12},
{\it solid} and {\it dashed curves} are computed 
(i.e. equations (\ref{eq:domdt}\,--\,\ref{eq:dadt}) are solved) starting with 
a subgiant model that has $R=3\,R_\odot$, assuming that initially $\Omega=0$, while
{\it dot-dashed} and {\it dotted curves} --- for
$\Omega = 2.5\times 10^{-5}$\,rad\,s$^{-1}$ (this value of $\Omega$
extrapolates to $v_{\rm rot}\approx 100$\,\kms\ back on the ZAMS);
{\it (ii)} from Fig.~\ref{fig:f13}, we can read the star's radius $R_{\rm cor}$ at which $\Omega/\omega = 0.9$; 
for the sake of simplicity, we assume that precisely at the moment when
$R=R_{\rm cor}$ the star's surface comes to corotation with the binary orbital motion;
we find that $R_{\rm cor}\approx 3.9\,R_\odot$ and $R_{\rm cor}\approx 6.7\,R_\odot$ 
for $a=50\,R_\odot$ and $a=80\,R_\odot$, respectively (Fig.~\ref{fig:f13});
{\it (iii)} now, we ``place'' our model star into a binary system; until the star's radius $R < R_{\rm cor}$, its evolution
is not considered to differ from the single star case, in particular, the whole star including its convective envelope is
assumed to rotate differentially with $j(t,\mr) = j(0,\mr)$; however, as soon as $R\geq R_{\rm cor}$, the solid body rotation 
$\Omega(r)=\omega$ of the outer part of the convective envelope, in which $R_{\rm cor}\leq r\leq R$, is enforced.

Our algorithm can create a faster rotating radiative zone in a red giant component of a wider binary 
in spite of the fact that a closer binary has a higher surface rotational velocity (compare
{\it solid} and {\it dotted} rotation profiles in Fig.~\ref{fig:f11}). Indeed, if in the case of
a tidally spun-up RGB star the specific angular momentum is still conserved in each mass shell 
with $R_{\rm c}\leq r \leq R_{\rm cor}$, then
the angular velocity in the radiative zone will depend on the radius as $\Omega(r) = \omega(R_{\rm cor}/r)^2$, where
\bea
\omega = 6.274\times 10^{-4}\left(\frac{\mstar}{\msol}\right)^{1/2}(1+q)^{1/2}\left(\frac{a}{R_\odot}\right)^{-3/2},\ \mbox{rad s}^{-1}.
\eea
Using this equation, for our solar metallicity red giant model in the binaries 
with $a=50\,R_\odot$ and $a=80\,R_\odot$ we find $\omega_1=0.2834$ ($R_{\rm cor}=3.9\,R_\odot$) and 
$\omega_2=0.1400$ ($R_{\rm cor}=6.7\,R_\odot$), respectively (both velocities are expressed in units of $10^{-5}$\,rad\,s$^{-1}$ 
and plotted in Fig.~\ref{fig:f10}). Hence, at the same location
in the radiative zone, the red giant in the binary with $a=80\,R_\odot$ rotates 1.5 times as fast as the one in the binary with $a=50\,R_\odot$,
in spite of the fact that $\omega_2 < \omega_1$.

\subsection{Possible Manifestations of Enhanced Extra Mixing}

\subsubsection{RS CVn binaries}

As primary constraints on possible manifestations of enhanced extra mixing in giant components of tidally locked binaries,
we have chosen the latest observational data on the CMD locations, rotational and orbital periods,
rotational velocities, and surface element abundances for a sample of late-type red giants in the RS CVn binaries
published by \cite{fea02}, \cite{mea04}, and \cite{fh05}. These red giants are chromospherically active stars whose activity is thought to be due
to their tidal spinning up that assists in generating strong magnetic fields via convective dynamo.
In order to be sure that all stars in the sample have achieved corotation, 
we select only those of them that have almost equal spin and orbital periods.
Thus selected objects are shown as {\it filled circles} in Fig.~\ref{fig:f9}. In the same figure, we have plotted evolutionary tracks
for single solar-metallicity stars with masses 1.3, 1.5, 1.7, 2, and $3\,\msol$ ({\it dashed curves}).
Looking at Fig.~\ref{fig:f9}, one cannot help but noticing that all 8 of the selected RS CVn binaries are located
on the lower RGBs. This striking photometric peculiarity has been emphasized by \cite{mea04}. To extend the size of the sample of
late-type red giants in tidally locked binaries, we have added a number of K-giants in binaries with circular orbits from Fig.~\ref{fig:f3}
to our selected RS CVn systems ({\it open circles} in Fig.~\ref{fig:f9}). We have used the Hipparcos parallaxes and Tycho magnitudes from
the online catalogue {\tt archive.ast.cam.ac.uk} to estimate the absolute magnitudes $M_V$ of the additional stars.
The data plotted in Fig.~\ref{fig:f9} show that the majority (15 out of 17) of the observed low-mass ($\mstar\la 2\,\msol$) binary stars, that are expected
to experience extra mixing on the RGB and that are most likely to have synchronized their spin and orbital rotation, 
are still located on the lower RGB. We will show that this photometric
peculiarity of the RS CVn and circularized binaries can be explained 
by rotational effects in the radiative zones of their tidally spun-up red giant components.

{\it Solid curve} in Fig.~\ref{fig:f9} is the evolutionary track of our solar-metallicity $1.7\,\msol$ model placed
into a binary system with $q=0.5$ and $a=80\,R_\odot$. As explained before, until the model's radius $R$ stays below     
the corotation radius $R_{\rm cor} = 6.7\,R_\odot$ (for $M_V\la 1.68$ on the RGB),
its evolution is not considered to differ from that of the single star.
In particular, the surface rotational velocity $v_{\rm rot}$ decreases in a way similar to that depicted by the dot-short-dashed curve
in Fig.~\ref{fig:f6}. However, as soon as $R$ exceeds $R_{\rm cor}$, the outer part of the model's convective envelope,
at $R_{\rm cor}\leq r\leq R$, is enforced to rotate like the solid body with the respective orbital velocity $\Omega(r)=\omega_2$ (see Fig.~\ref{fig:f10}).
It means that from the moment when $R$ becomes equal $R_{\rm cor}$ on the RGB ($M_V\approx 1.68$) the surface rotational velocity $v_{\rm rot}$ of
our model starts to increase following the thin solid curve in Fig.~\ref{fig:f12}. After this moment, while the surface angular velocity of
our binary model's single counterpart continues to decrease due to the conservation of the total angular momentum and
the red giant's expansion, the outer convective layers of our binary model keep rotating at the same angular velocity
$\omega_2$ and its surface rotational velocity $v_{\rm rot}=R\,\omega_2$ increases with $R$. 
By the bump luminosity, the ratio of $\omega_2$ to $\Omega(R)$ for the single model grows up to $\sim$\,10 (compare
the dot-dashed and solid curves at the maximum $r$ in Fig.~\ref{fig:f11}). 
At the BCE, this ratio is somewhat smaller because in the convective envelope of
the single model $\Omega(r)$ increases with decreasing $r$ all the way from $R$ to $R_{\rm BCE}$, while in the binary model
$\Omega(r)$ is maintained equal to $\omega_2$ from the surface down to the depth $R_{\rm cor}$. Beneath the BCE, the same difference
between the angular velocities for the binary and single star case that has been settled at the BCE is slowly propagating through the radiative
zone, like a wave, during the subsequent stellar evolution (see the solid and short-dashed curves in Fig.~\ref{fig:f11}).

Usually, when the HBS crosses the chemical composition discontinuity left behind by the BCE at the end of the first dredge-up,
a low-mass star makes a tiny zigzag on the CMD, like that outlined by a small square in Fig.~\ref{fig:f4}.
However, in our case of the solar-metallicity $1.7\,\msol$ binary star the zigzag is found to be unusually big (the solid
curve between $M_V\approx 0.7$ and $M_V\approx 1.5$ in Fig.~\ref{fig:f9}). This is a new result which is entirely due to rotational effects.
Indeed, in panel a in Fig.~\ref{fig:f14} the corotation radius is varied from 2 to $8\,R_\odot$, thus accelerating rotation of the
radiative zone in the bump luminosity model of our $1.7\,\msol$ binary star, while panel b demonstrates how 
the zigzag's extent responds to the variations of $R_{\rm cor}$. These test computations have been done without extra mixing,
just with rotational effects on the stellar structure and evolution included according to \cite{dv03b}. 
The region in the CMD in which our binary star makes the zigzag is comparable by its size with the area occupied by
the majority of the observed tidally locked binaries (Fig.~\ref{fig:f9}). Therefore, we surmise that it is the extended bump luminosity zigzags
produced by tidal spinning up of the convective envelopes and underlying
radiative zones that are responsible for the fact that the late-type giant components of the synchronized
binaries predominantly reside on the lower RGB. This hypothesis receives a further support if we compare the evolutionary times
or differential luminosity functions of the single and binary model stars in the vicinity of the bump luminosity.
Whereas the time spent by the single star ({\it thick curve} in Fig.~\ref{fig:f15}) between $M_V=2$ and
$M_V=0$ is 109 million years, it takes 146 million years for the binary star ({\it thin curve} in Fig.~\ref{fig:f15}) 
to make the zigzag on the CMD when its magnitude $M_V$ first increases from 0.7 to 1.5 and then decreases back to 0.7.
Besides, the bump luminosity itself is shifted by $\Delta M_V\approx 0.8$ towards the subgiant branch in the binary star case, which    
might explain the fact that the low-mass red giants in tidally locked binaries ``prefer'' to reside on the lower RGB.

For 7 out of our 8 selected RS CVn binaries, measured values of $v\sin i$ are available in the cited papers.
4 binaries have orbital periods close to 50 days, for the rest 3 of them $P\approx 25$ days.
In most cases, an estimated mass of the primary component lies between 1.5 and $1.7\,\msol$, and its metallicity $-0.09\leq$\,[Fe/H]\,$\leq 0.12$. 
That is why we have chosen the solar-metallicity $1.7\,\msol$ model star placed in binaries with $a=50\,R_\odot$ ($P_{\rm d}\approx 25$)
and $a=80\,R_\odot$ ($P_{\rm d}\approx 50$) as one of our basic binary models. 
Note that for $P_{\rm d} = 25$ and $P_{\rm d} = 50$, our binary $1.7\,\msol$ model with $q=0.5$ will start to transfer
matter to its companion by Roche-lobe overflow at $M_V\approx -0.03$ and $M_V\approx -0.65$, respectively,
i.e. well above the observed location of the RS CVn binaries in the CMD (Fig.~\ref{fig:f9}). Therefore, their scarcity on upper RGBs
can hardly be attributed to a dissipation of their envelopes on a short (dynamical) timescale in a common envelope event
that follows the Roche-lobe overflow.
Fig.~\ref{fig:f12} shows that, within observational errors, our solutions
of equations (\ref{eq:domdt}\,--\,\ref{eq:dadt}), that control the tidal evolution of the red giant's angular velocity
and the binary's semi-major axis, conform with the observed locations of the selected RS CVn binaries on the $M_V$ versus $v\sin i$ plane.

The tidal spin-up of the radiative zone may transform extra mixing from its canonical to an enhanced mode, with
the mixing depth and rate increased proportionally to $\Omega^2$ according to equations (\ref{eq:dvmix}\,--\,\ref{eq:dvmmix}).
{\it Dot-dashed curves} in panels b and d in Fig.~\ref{fig:f16} demonstrate how the diffusion coefficient
$D_{\rm v}$ from equation (\ref{eq:dv}) with $f_{\rm v}=1$
changes when the envelope material spun up by the tidal force eventually arrives at
the HBS in our 1.7\,$\msol$ upper RGB\footnote{We still call ``the upper RGB models'' those spun-up binary red giants
in which the HBS has erased the composition discontinuity left behind by the BCE at the end of the first dredge-up in spite of
the fact that they can reside on the lower RGB for quite a long time thereafter.} binary model. 
Comparing panels c and d, we find that
the tidally enforced enhanced extra mixing can dredge up some fresh Na that is being synthesized from $^{22}$Ne in the HBS.
This finding is not surprising because it has already been predicted for spun-up upper RGB stars by \cite{dv03a}.
What is really surprising is that Na overabundances of the right magnitude have recently been reported in 5 out of our 8 selected
RS CVn binaries by \cite{mea04}. These are plotted in Fig.~\ref{fig:f17} against the Li abundances measured by the same authors.
They have applied NLTE corrections to both [Na/Fe] and $\eli$. They have also claimed that ``chromospheric
heating appears insufficient to account for the anomalously high Na abundances''.

We have included the effect of tidal spin-up on $D_{\rm v}$ in our stellar evolution computations.
The resulting enhanced extra mixing produces {\it dashed curve} in Fig.~\ref{fig:f17}. This curve 
corresponds to the stellar evolution that starts
at the bump luminosity, includes the rotationally extended zigzag along the lower RGB and ends well above the bump luminosity.
However, since most of the tidally synchronized binary red giants (including all 8 of our selected RS CVn binaries)
in Fig.~\ref{fig:f9} are still located in the zigzag region, only the part of the dashed curve that corresponds
to the evolution before the end of the zigzag should be compared with the observational
data points (a region between {\it dotted line segments} in Fig.~\ref{fig:f17}). 
By the end of this part, [Na/Fe] reaches 0.36 and $\eli$ accumulates only -0.6
after its initial drop by more than 3 orders of magnitude.
These values are smaller than the observed abundances. 

A much better result ({\it solid curve} in Fig.~\ref{fig:f17}) is obtained with
the constant diffusion coefficient $\dm = 10^{11}$\,\cs\ ({\it dotted curve} in panel d in Fig.~\ref{fig:f16}).
In this case, we have actually made three simplifying assumptions: 
{\it (i)} that canonical extra mixing has operated with the constant diffusion 
coefficient $\dm = 10^9$\,\cs\ ({\it dotted curve} in panel b in Fig.~\ref{fig:f16}), which is close to
the semi-empirical values estimated by \cite{dv03a}; {\it (ii)} the tidal synchronization of our binary red giant model
has increased $\Omega$ in its radiative zone exactly by the factor of 10, 
which is even slightly less than the ratio between the $\Omega$-profiles
for the binary and single star case seen in Fig.~\ref{fig:f11}; {\it (iii)} like $\dm$ from equation (\ref{eq:dvmix}),
our uniform diffusion coefficient scales as $\dm\propto\Omega^2$. It is important to note that,
notwithstanding these approximations, the mixing depth has yet been determined
by equation (\ref{eq:dvmmix}) because we believe firmly that it is constrained by the kinetic energy available from
differential rotation of the radiative zone. 

The difference between the dashed and solid curve in Fig.~\ref{fig:f17}
is mainly due to the behaviour of the Li abundance, which can be explained by the same reason why $\eli$ changes 
differently in a single star on the upper RGB towards the RGB tip in the secular shear instability and simple diffusion
model of canonical extra mixing (compare upper panels in our Fig.~\ref{fig:f2} and in Fig.~4 from \cite{dv03a}). In the first case,
$\dm$ is proportional to the thermal diffusivity $K$ (equation (\ref{eq:dvmix})) that strongly decreases with the luminosity
when our tidally spun-up red giant descends along the extended bump luminosity zigzag ({\it solid curve} in Fig.~\ref{fig:f9}).
Thus, on the one hand, the tidal spinning up tends to enhance $\dm$ because $\dm\propto\Omega^2$ but, on the other hand,
the effect of fast rotation on the red giant's internal structure causes its luminosity, and hence the value of $K$ in the radiative zone,
to drop, so the resulting mixing rate $\dm\propto\Omega^2K$ turns out to be too slow, compared to
the case of constant $\dm = 10^{11}$\,\cs, to produce and support (via the $^7$Be-mechanism)
as high Li abundance as in the RS CVn binaries (Fig.~\ref{fig:f17}). Of course, if, in concordance with their 
apparent residence on the lower RGB (Fig.~\ref{fig:f9}), the RS CVn binaries have not reached their bump luminosities yet
then their surface Li abundances are in good agreement with those predicted by the standard theory for
the first dredge-up ({\it solid curve} at [Na/Fe]\,$< 0.05$ in Fig.~\ref{fig:f17}). However, in that case, RS CVn binaries
would not have had as large Na overabundances as those reported by \cite{mea04}. A spectroscopic determination of CN abundances
in the giant components of the RS CVn binaries could verify our hypothesis of enhanced extra mixing in these stars. Indeed,
if all the assumptions made are correct then, according to our computations, 
these stars should have \cc\,$\approx 3.9$, [C/Fe]\,$\approx -1.8$ and [N/Fe]\,$\approx 0.63$
instead of the standard post-first-dredge-up values of \cc\,$\approx 25$, [C/Fe]\,$\approx -0.17$ and [N/Fe]\,$\approx 0.30$.

\subsubsection{Binary Red Giants in Globular Clusters}

The problem of Na overabundances in RS CVn binaries and its possible solution proposed by us 
may be related to the long-standing problem of star-to-star abundance variations of C, N, O, Na, Mg and Al in globular clusters.
At present, there are no doubts that a sizeable component of those variations originated from hydrostatic hydrogen burning in more massive stars
that had completed their lives in the past. Indeed, firstly, there are correlations between overabundances of
N, Na and Al and deficits of C, O and Mg as if all of them resulted from
simultaneous operation of the CNO, NeNa and MgAl cycle. Secondly, the same abundance variations that are seen in red giants are also found 
in subgiant and MS stars, the latter evidently having too low central temperatures to produce
these variations {\it in situ}. Hence, some of the globular-cluster MS stars 
must have accreted large amounts of material (or they have entirely been formed of it) processed in H-burning that had been
ejected by now extinct stars into the interstellar medium of globular clusters.

Figs.~\ref{fig:f18} and \ref{fig:f19} give nice examples of the O--Na anticorrelation and
Na--Al correlation for large samples of red giants in the globular clusters M3 ({\it circles}) and M13 ({\it triangles})
recently updated by \cite{sea04}. If we take [O/Fe]$_{\rm init} = 0.3$, [Na/Fe]$_{\rm init} = -0.2$ and 
[Al/Fe]$_{\rm init} = -0.2$ as the initial abundances in those M3 and M13 MS stars that were not polluted by
their more massive cluster-mates, and [O/Fe]$_{\rm acc} = -1.0$, [Na/Fe]$_{\rm acc} = 0.7$ and 
[Al/Fe]$_{\rm acc} = 1.2$ as the most extreme abundances in the material
accreted by some of the cluster MS stars then both correlations can be considered as simple mixtures of
a $(1-x)$-fraction of the material with the initial abundances and an $x$-fraction of the (accreted) material with the
most extreme abundances ({\it solid curves} in Figs.~\ref{fig:f18} and \ref{fig:f19}; {\it thick dots} on the curves
mark the values of $x$ from 0 to 1 with the increment 0.1). So, the next question is what H-burning stars
were capable of producing those extreme abundances of O, Na and Al. There are two alternative answers to this question:
(1) intermediate-mass (4\,--\,6\,$\msol$) asymptotic giant branch (AGB) stars (e.g. \citealt{vda05}), or (2)
upper RGB stars that had been more massive (say, $0.9<\mstar/\msol\la 2$) than the present-day globular-cluster MS turn-off stars
and that had experienced enhanced extra mixing caused by their tidal spin-up in binary systems (\citealt{dv03a,dw04}).

Since the first alternative is not without its shortcomings (e.g., when destroying O in intermediate-mass
AGB stars the (hydrogen) hot-bottom burning depletes $^{24}$Mg even stronger (\citealt{dh03}); 
it also keeps [C/Fe]\,$\ga -0.5$ (\citealt{dw04}); besides, the third dredge-up in AGB stars
should increase the total abundance of C+N+O (\citealt{fea04}); {\it none} of these theoretical predictions is supported by
observations),
the second idea is worth developing. Pursuing this goal, we have computed the evolution of
the $1\,\msol$ star with the helium mass fraction $Y=0.24$ and $Z=0.0005$, which gives a metallicity [Fe/H] close
to those of both M3 ([Fe/H]\,=\,$-1.57$) and M13 ([Fe/H]\,=\,$-1.54$). In order to simulate tidally
enforced enhanced extra mixing, this model star has been placed into a binary system
with $q=0.3$ and $a=50\,R_\odot$. We used the same computational method and approximations as in the case of RS CVn binaries.
The initial rotation parameter was $f_\varepsilon = 0.0003$,
as in our $0.85\,\msol$ single star model with $Z=0.0005$,
and the enhancement factor in equation (\ref{eq:dvmix}) was $f_{\rm v} = 20$.
The computed evolution of the O and Na surface abundances is shown in Fig.~\ref{fig:f18} with {\it dashed curve}.
In order to produce enough Al ({\it dashed curve} in Fig.~\ref{fig:f19}) we had to make the same additional assumptions
as in \cite{dt00}, namely: (1) the initial abundance of the $^{25}$Mg isotope was increased to [$^{25}$Mg/Fe]$_{\rm init} = 1.2$
compared to the scaled solar ratio, and (2) the reaction $^{26}$Al$^{\rm g}$(p,$\gamma)^{27}$Si was sped up
by the factor of $10^3$ compared to its rate from \cite{aea99}. 

Comparing the binary red giant evolution with
that of a single red giant with the same mass and composition, we have found that the tidal spin-up has accelerated
rotation of the radiative zone by a factor of $\sim$\,20. Taking into account intrinsic imperfections of our description of
the tidal spin-up, we have also considered a test case in which rotation of the red giant's radiative zone has 
artificially been maintained 25 times as fast as in the single red giant model. For this test case, the correlated O, Na and Al
abundance variations are depicted with {\it dotted curves} in Figs.~\ref{fig:f18} and \ref{fig:f19}.
In both cases, the abundances of O, Na and Al naturally evolve from their initial
values of [O/Fe]$_{\rm init} = 0.3$, [Na/Fe]$_{\rm init} = -0.2$ and [Al/Fe]$_{\rm init} = -0.2$ 
close to the most extreme values supposedly present in the material accreted by globular-cluster MS stars.

It is important to note that environmental conditions in globular clusters are likely to be appropriate for our binary star
scenario to contribute to the star-to-star abundance variations. Firstly, frequent single-binary and binary-binary stellar
encounters work toward decreasing the semi-major axes of hard binaries (\citealt{hea92}). In particular, \cite{bd04}
have recently shown that after $\sim$\,20 of such encounters the initial $\log a$-flat binary population transforms
into a gaussian-like distribution with a peak at $a\approx 100\,R_\odot$. Secondly, binary red giants with $a\leq 100\,R_\odot$
will definitely fill their Roche lobes before reaching the RGB tip. Therefore they will probably lose their
convective envelopes very quickly during a common-envelope phase (\citealt{bd04}). According to our hypothesis,
these envelopes will be enriched with Na and Al while O will be depleted due to enhanced extra mixing triggered by
their tidal spin-up. Finally, a fraction of primordial binaries in globular clusters migh have been very high,
up to 100\% (\citealt{iea05}).

\section{Concluding Remarks}
\label{sec:concl}

In this paper, we have elaborated upon the ideas proposed by \cite{dv03a} about canonical extra mixing in single upper RGB stars
and enhanced extra mixing in low-mass red giants spun-up as a result of their tidal synchronization
in close binary systems. In order to put as many as possible observational constraints on properties of extra mixing,
we have supplemented the old data on the Li and CN abundance
changes along upper RGBs in field low-metallicity stars with the new data on photometry, chemical peculiarities
and rotational periods/velocities of stars from the solar-metallicity open cluster M67, globular clusters
47\,Tuc, NGC\,6397, NGC\,6752, and RS CVn binaries.

We have confirmed the conclusions made in the earlier paper that
the secular shear instability driven by differential rotation of the red giants' radiative zones can be considered
as a promising physical mechanism for both modes of extra mixing while the turbulent diffusion coefficient (\ref{eq:dv}) derived
by \cite{mm96} can be used to model them appropriately, provided that: {\it (i)} unlike the Sun, low-mass MS progenitors of those red giants
already possessed differentially rotating radiative cores; {\it (ii)} the specific angular momentum was conserved in each mass shell,
including convective regions, inside those stars during their entire evolution from the MS through the RGB tip; and
{\it (iii)} the diffusion coefficient (\ref{eq:dv}) is taken with the enhancement factor $f_{\rm v}\approx 20$.
For a binary red giant, we have additionally assumed that the tidal force brings to corotation only an upper part of
its convective envelope. For the orbital and stellar parameters typical for the RS CVn binaries, this assumption results
in spinning up of their radiative zones by a factor of $\ga$\,10. 

Although we present some arguments in support of
the made assumptions, it still seems unlikely that even our tidally enforced enhanced extra mixing with $\dm\sim 10^{11}$\,\cs\ 
is not accompanied by a fast transport of angular momentum that would work toward flattening the $\Omega$-profile in the radiative zone,
thus reducing the mixing efficiency. From this point of view, our approach to modeling extra mixing in upper RGB stars
is similar to the ``Maximal Mixing Approach'' recently used by \cite{chea05}. The question ``how far is our approach from reality?''
can be answered in the usual way --- by comparing the theoretical predictions made by us with future observations.
Here are some possible observational tests that can support or reject our models. 

First, we propose to determine \cc\ ratios
in upper RGB stars in the solar-metallicity open cluster NGC\,6791 whose MS turn-off mass is $\sim\,1.1\,\msol$.
If the Sun is not an exceptional case then, like the Sun, MS progenitors of the NGC\,6791 red giants must have rotated as solid bodies.
Therefore, we expect that canonical extra mixing in upper RGB stars in NGC\,6791 should be inefficient. Hence, these stars
should keep their post-first-dredge-up values of \cc\,$\approx 25$ unchanged instead of having the values of \cc\,$\approx 10$
similar to those measured in the M67 evolved stars that have definitely experienced canonical extra mixing.

\begin{figure}
\plotone{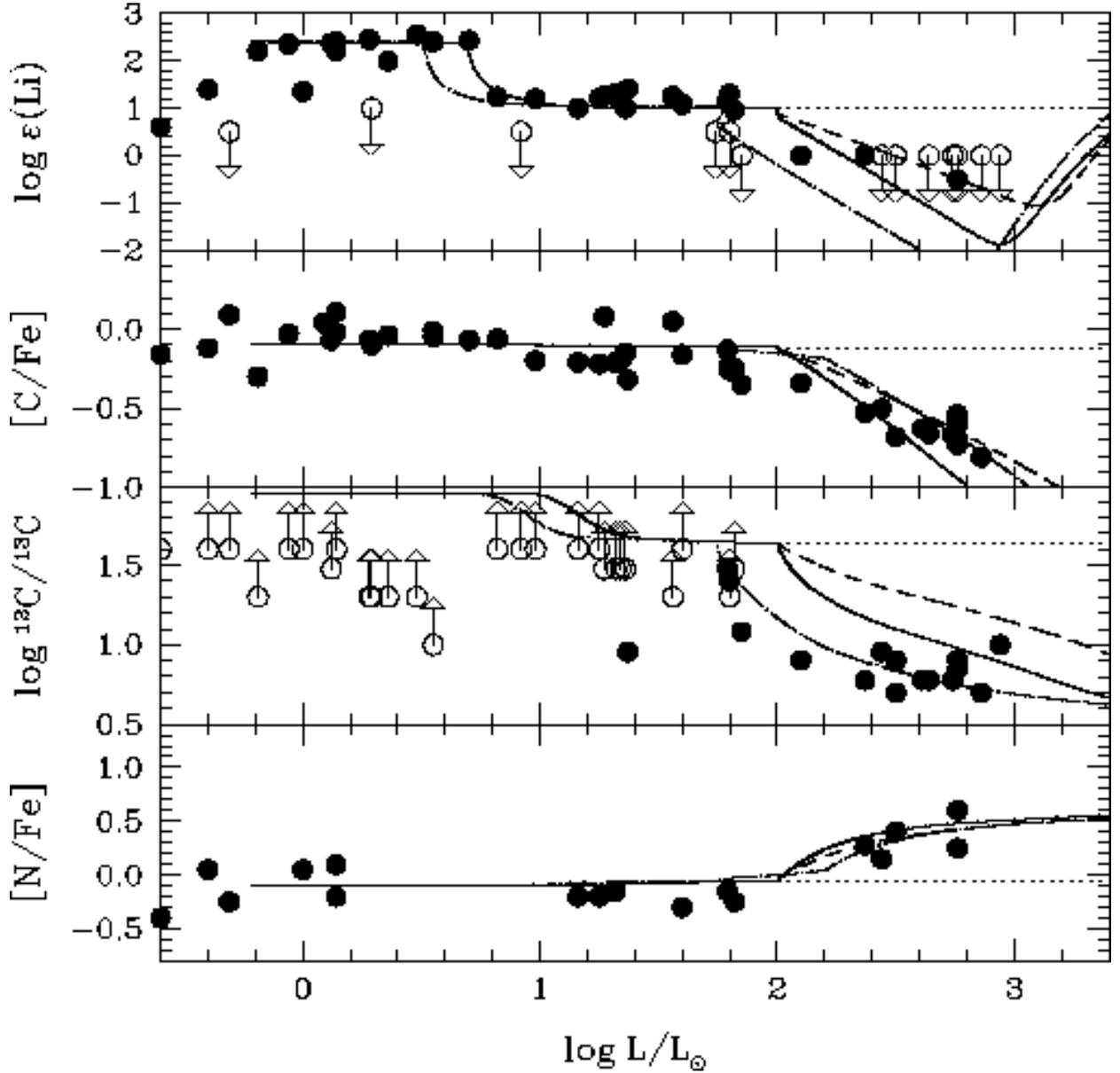}
\caption{Comparison of the observational data from \cite{grea00}
         for field metal-poor ($-2\la\mbox{[Fe/H]}\la -1$) low-mass stars
         ({\it circles}) with results of our stellar evolution
         computations with rotational mixing on the upper
         RGB described by equation (\ref{eq:dv}).
         It is assumed that the specific angular momentum is conserved
         in each mass shell, including convective regions, during
         the entire stellar evolution. Computations are done for
         $\mstar = 0.85\msun$ and the initial rotation parameter
         $f_\varepsilon = 0.0003$ (equation (\ref{eq:initrot})). 
         Theoretical results are presented for the following
         combinations of the enhancement factor
         ($f_{\rm v}$ in equation (\ref{eq:dv})) and heavy-element mass
         fraction: $f_{\rm v}=25$ for $Z=0.0005$ and $Z=0.002$
         ({\it solid and dot-dashed curves}, respectively), and
         $f_{\rm v}=15$ for $Z=0.0005$ ({\it dashed curve}).
        }
\label{fig:f2}
\end{figure}

\begin{figure}
\plotone{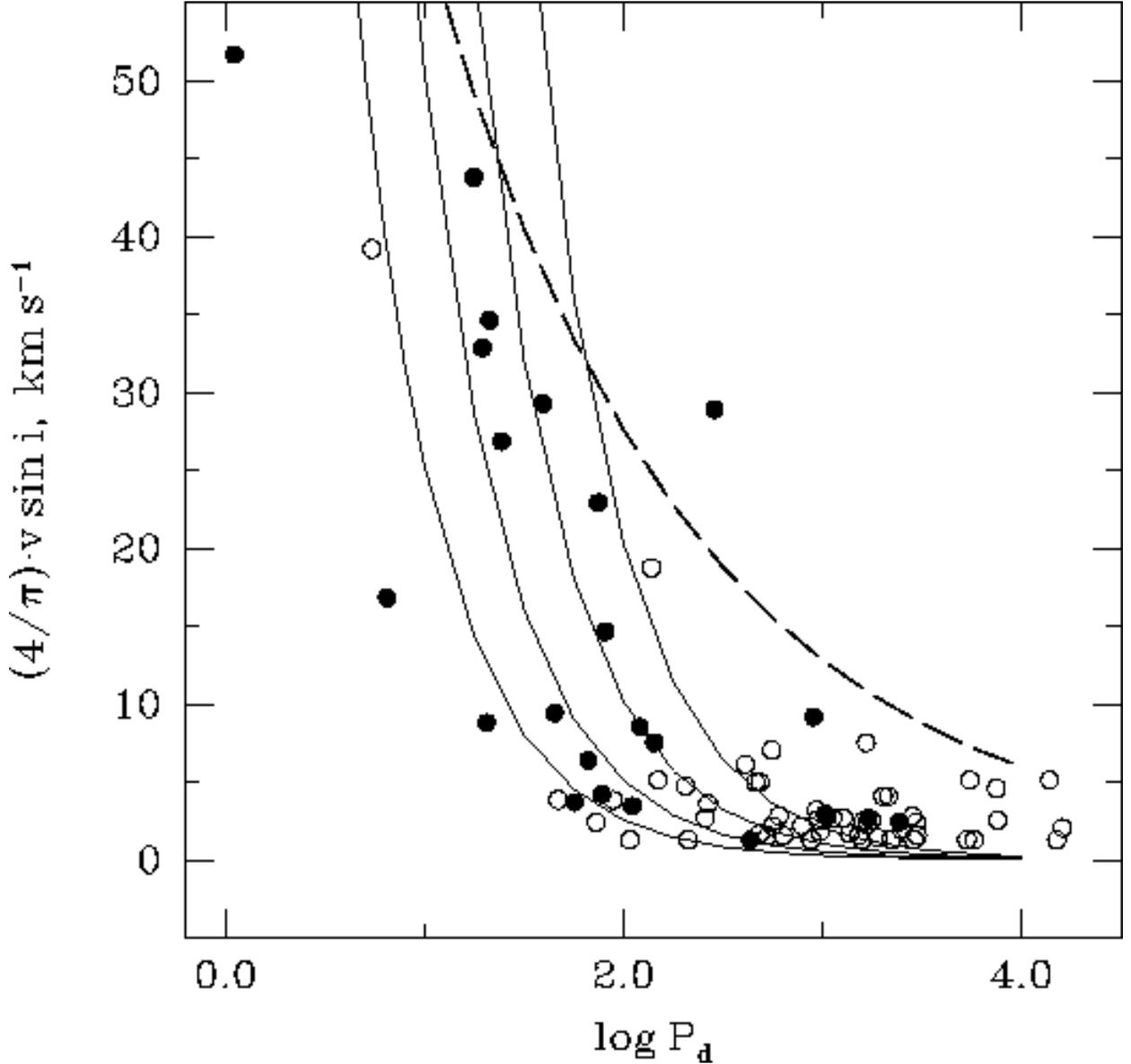}
\caption{An illustration of tidal synchronization in real stars. {\it Circles} are
         G and K giant components of field binaries observed by \cite{dmea02}.
         The factor $4/\pi\approx 1.27$ takes into account the random orientation of
         their rotation axes. $P_{\rm d}$ is the binary orbital period in days.
         {\it Filled circles} are binaries with nearly circular
         orbits (eccentricities $e\leq 0.10$). The systems with $e > 0.10$ are represented by {\it open circles}.
         A set of theoretical {\it solid curves} is constructed by us assuming that $\Omega = \omega$, where $\Omega$ and $\omega$
         are the spin and orbital angular velocity of a red giant, for radii $R/R_\odot =$\,5, 10, 20 and 40
         (from left to right). The giants represented by filled circles are most likely to have synchronized their rotation.
         {\it Dashed curve} shows the maximum possible $v_{\rm rot}$ of a red giant that fills its Roche
         lobe in a binary system with $\mstar = 1.7\,\msol$, $q=0.5$.
        }
\label{fig:f3}
\end{figure}

\begin{figure}
\plotone{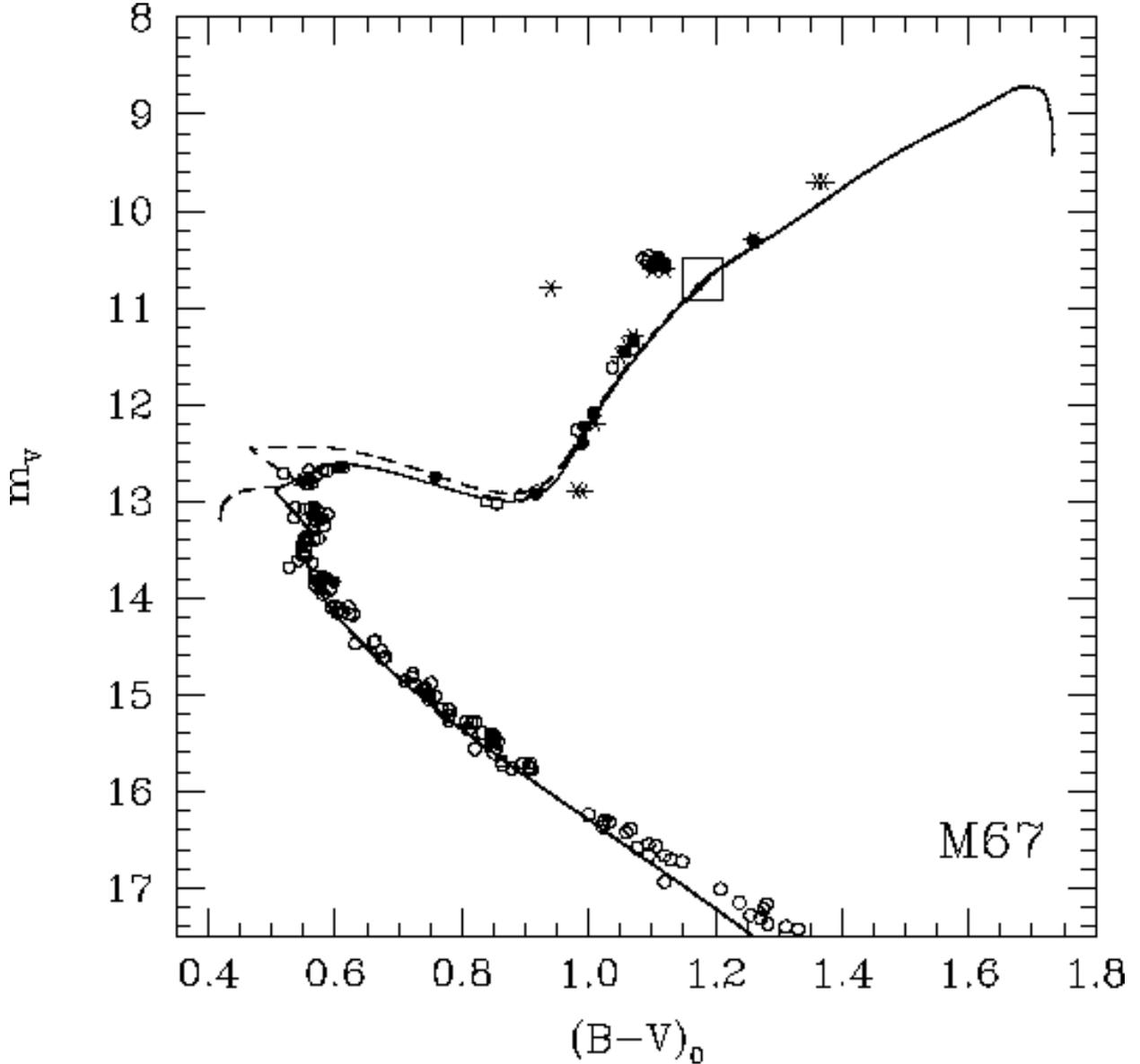}
\caption{High precision CMD of the old solar-metallicity open cluster M67
         ({\it open circles}) from the data of \cite{s04} and our $4\times 10^9$
         yr isochrone ({\it solid curve}). A region around
         the bump luminosity is outlined by a square. {\it Dashed        
         curve} is the evolutionary track of the model star with
         $\mstar = 1.35\,\msol$ and $Z=Z_\odot =0.0188$. All models were computed with the same
         initial rotation parameter $f_\varepsilon = 0.00075$.
         {\it Filled circles} are stars with known values of $v\sin\,i$
         from \cite{mea01}. {\it Asterisks} are red giants and clump stars for which the ratio
         \cc\ was measured by \cite{gb91}. For the cluster reddening and distance modulus,
         we have adopted the values $E(B-V)=0.03$ and $(m-M)_V=9.67$ that correspond to
         the observational lower limits provided by \cite{s04}.
        }
\label{fig:f4}
\end{figure}

\begin{figure}
\plotone{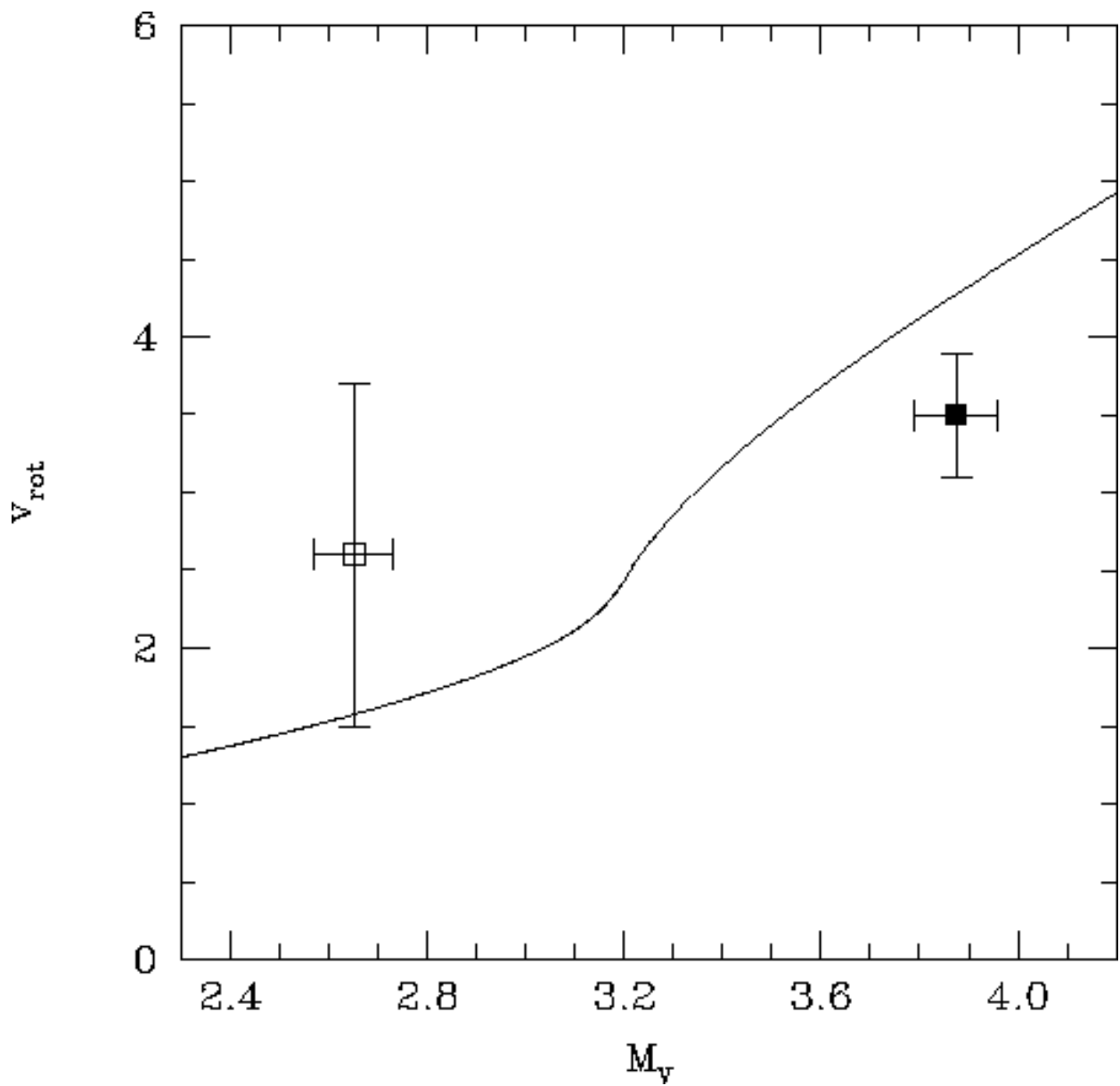}
\caption{Estimates of the upper limits for the ``true'' mean rotational
         velocities of stars in globular clusters 47~Tuc, NGC\,6397 and
         NGC\,6752 by \cite{lg03}. {\it Filled square} --- a mean for MS turn-off
         stars, {\it open square} --- for subgiants. Horizontal errorbars embrace
         all $M_V$ values for the observed samples of stars in NGC\,6752 ([Fe/H]\,=\,$-1.43$).
         Theoretical curve shows how $v_{\rm rot}$ changes during the evolution of
         our model star  with $\mstar = 0.85\,\msol$ and $Z=0.0005$ in which
         the specific angular momentum is conserved in each mass shell.
        }
\label{fig:f5}
\end{figure}

\begin{figure}
\plotone{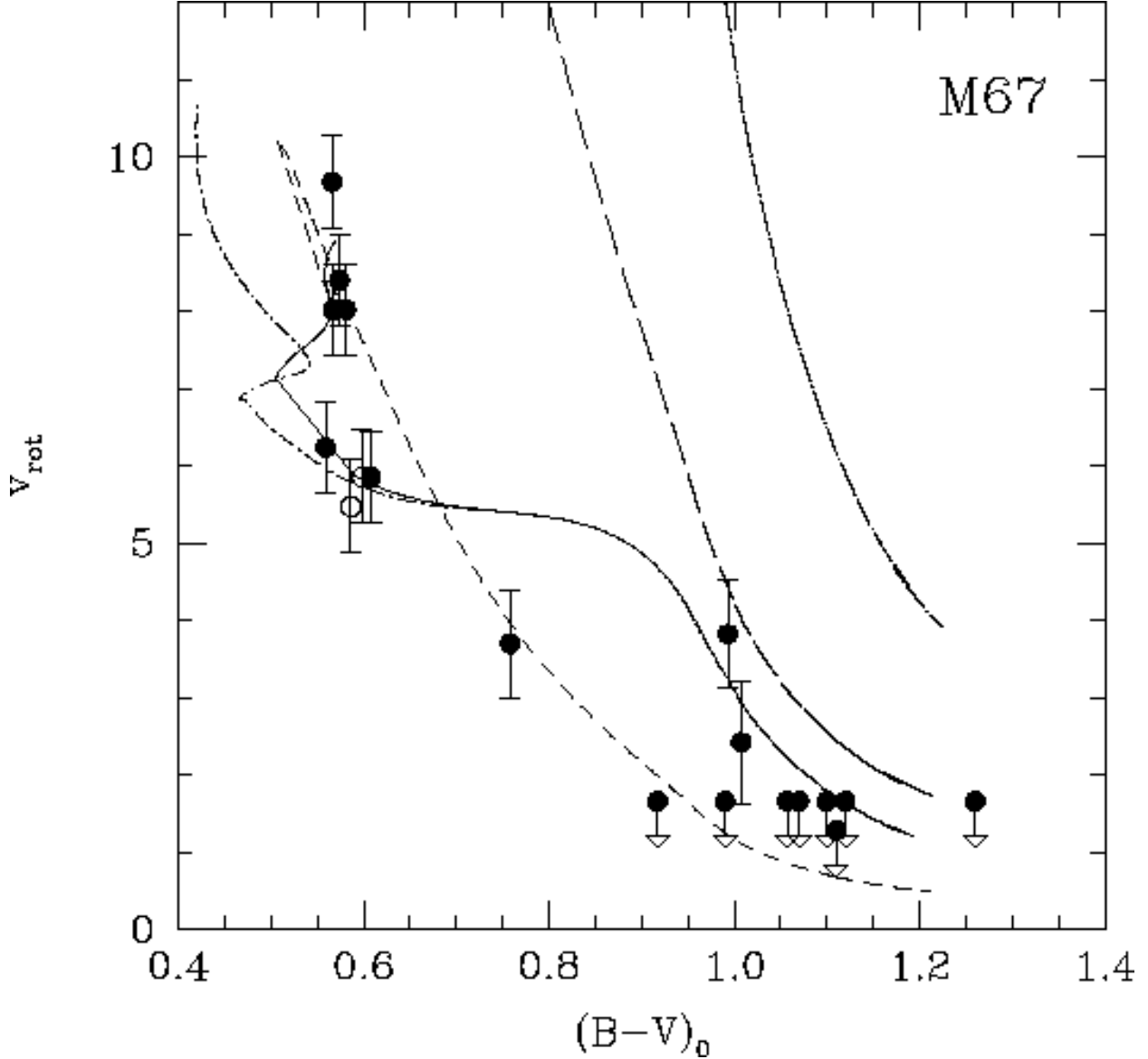}
\caption{Observational data on $(4/\pi)\,v\sin\,i$ for M67 stars
         ({\it open circles} are MS stars, {\it filled circles} are
         MS turn-off, subgiant and lower RGB stars) from \cite{mea01} and theoretical
         dependences of $v_{\rm rot}$ on $(B-V)_0$. {\it Solid curve} ---
         the $4\times 10^9$ yr rotational isochrone for $Z=0.0188$.
         {\it Dot-short-dashed curve} shows the rotational evolution of our $1.35\,\msol$
         model. Both curves are computed under the assumption of constant specific
         angular momentum in each stellar mass shell for the initial rotation parameter $f_\varepsilon = 0.00075$.
         {\it Short-dashed curve} is the $4\times 10^9$ yr isochrone
         with the initial velocity $v_{\rm rot}\approx 10$\,\kms\ 
         computed assuming solid-body rotation inside all of the $\mstar\geq 1.1\,\msol$ models
         used to construct it. For comparison,
         {\it dot-long-dashed curve} and {\it long-dashed curve} represent the evolution of
         the $1.35\,\msol$ model for the cases of differential and solid-body rotation,
         respectively, but for the initial velocity $v_{\rm rot}\approx 40$\,\kms.
        }
\label{fig:f6}
\end{figure}

\begin{figure}
\plotone{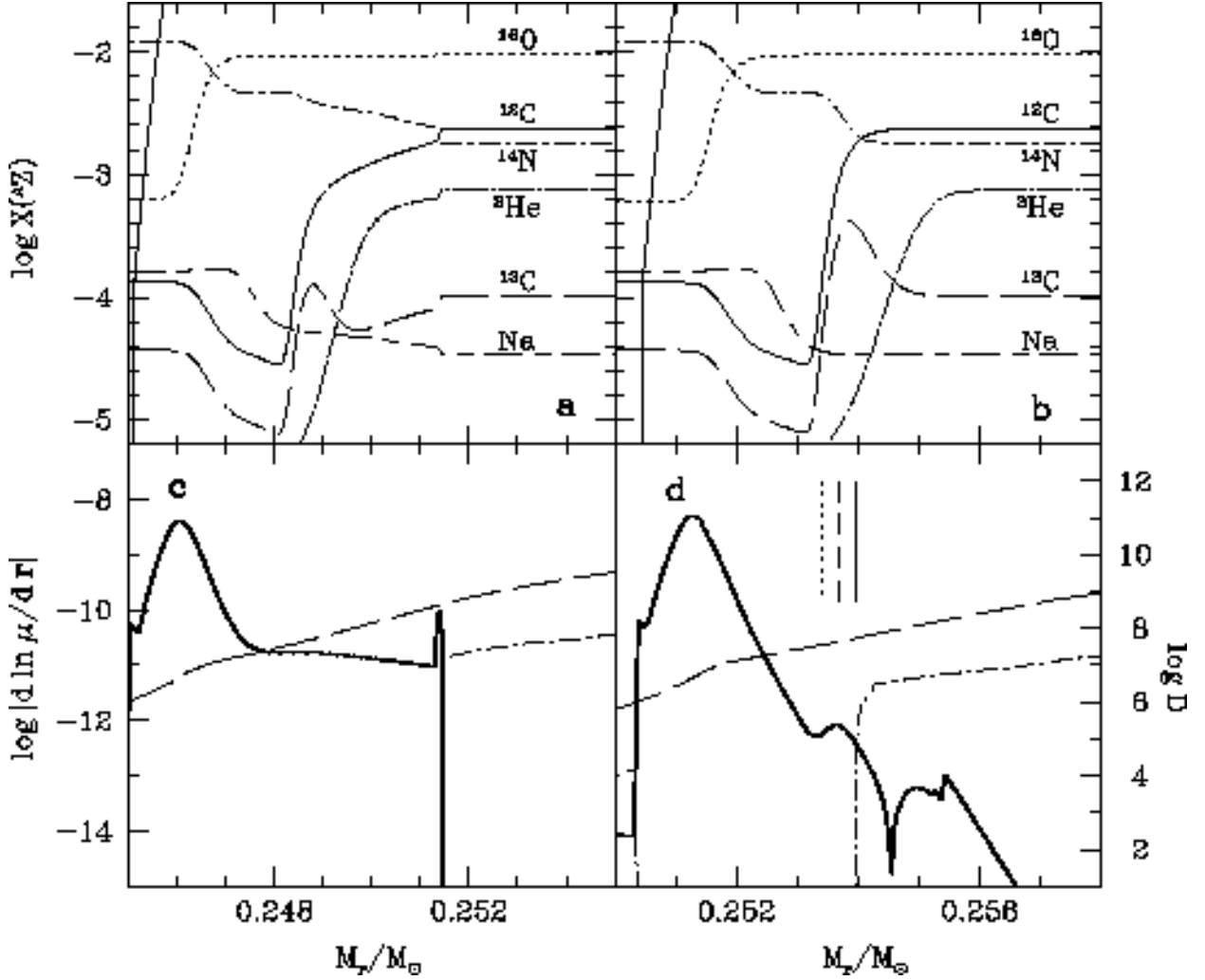}
\caption{Profiles of some element abundances (mass fractions in panels a and b),
         $\mu$-gradients and diffusivities (panels c and d) near the HBS
         in two rotating RGB models with $\mstar = 1.35\,\msol$, $Z = 0.0188$ and
         the initial rotation parameter $f_\varepsilon = 0.00075$. The models are located immediately
         above (panels b and d) and below (panels a and c) the bump luminosity.
         In panels c and d, the $\mu$-gradients are plotted using {\it solid curves},
         thermal diffusivity $K$ --- {\it long-dashed curves}, coefficients of
         the vertical turbulent diffusion $D_{\rm v}$ 
         (these are given by equation (\ref{eq:dv}) for the enhancement factor $f_{\rm v}=15$)
         --- {\it dot-dashed curves}. Vertical line segments in panel d
         point to the extra mixing depths specified by parameters $\dlt = 0.19$ ({\it dotted line})
         and $\dlt = 0.22$ ({\it dashed line}) as well as to the depth determined by
         equation (\ref{eq:dvmmix}) ({\it solid line}).
        }
\label{fig:f7}
\end{figure}

\begin{figure}
\plotone{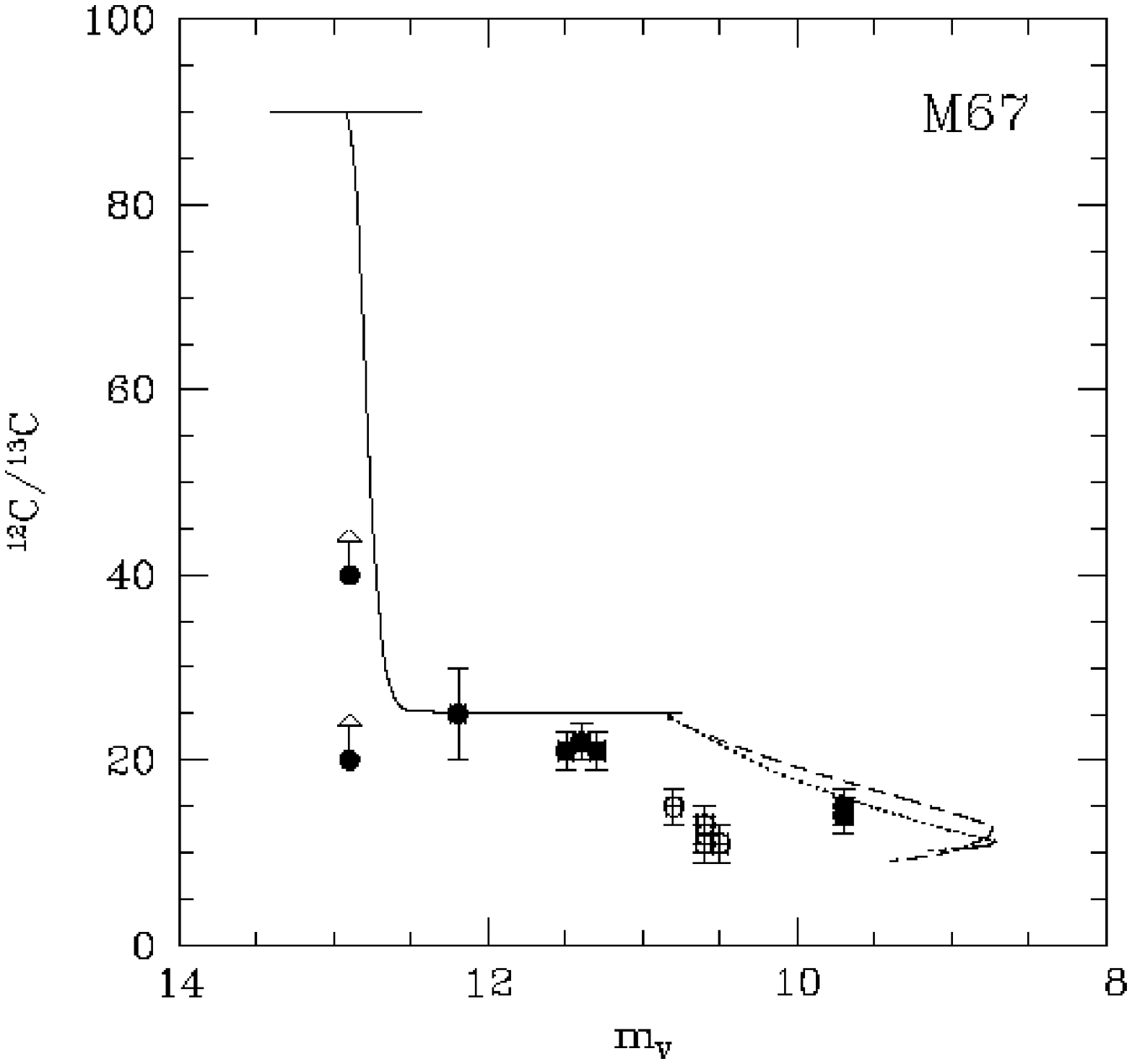}
\caption{\cc\ isotopic ratios in the M67 evolved stars measured by \cite{gb91}
         ({\it symbols with arrows and errorbars}). {\it Filled circles} are RGB stars
         while {\it open circles} are clump stars. {\it Solid curve} shows a decrease
         of \cc\ produced by the standard first dredge-up for the model star with
         $\mstar = 1.35\msun$ and $Z = 0.0188$. {\it Dotted curve} presents results of
         our computations in which extra mixing on the upper RGB has been simulated using
         the diffusion model with the depth and rate parameters:
         $\dlt = 0.22$ and $\dm = 8\times 10^8$\,\cs. {\it Dashed curve} is obtained
         for our rotating evolutionary model with extra mixing modeled by the vertical
         turbulent diffusion (the diffusion coefficient from equation (\ref{eq:dv}) with
         the enhancement factor $f_{\rm v}=15$). The initial rotation parameter was $f_\varepsilon = 0.00075$.
        }
\label{fig:f8}
\end{figure}

\begin{figure}
\plotone{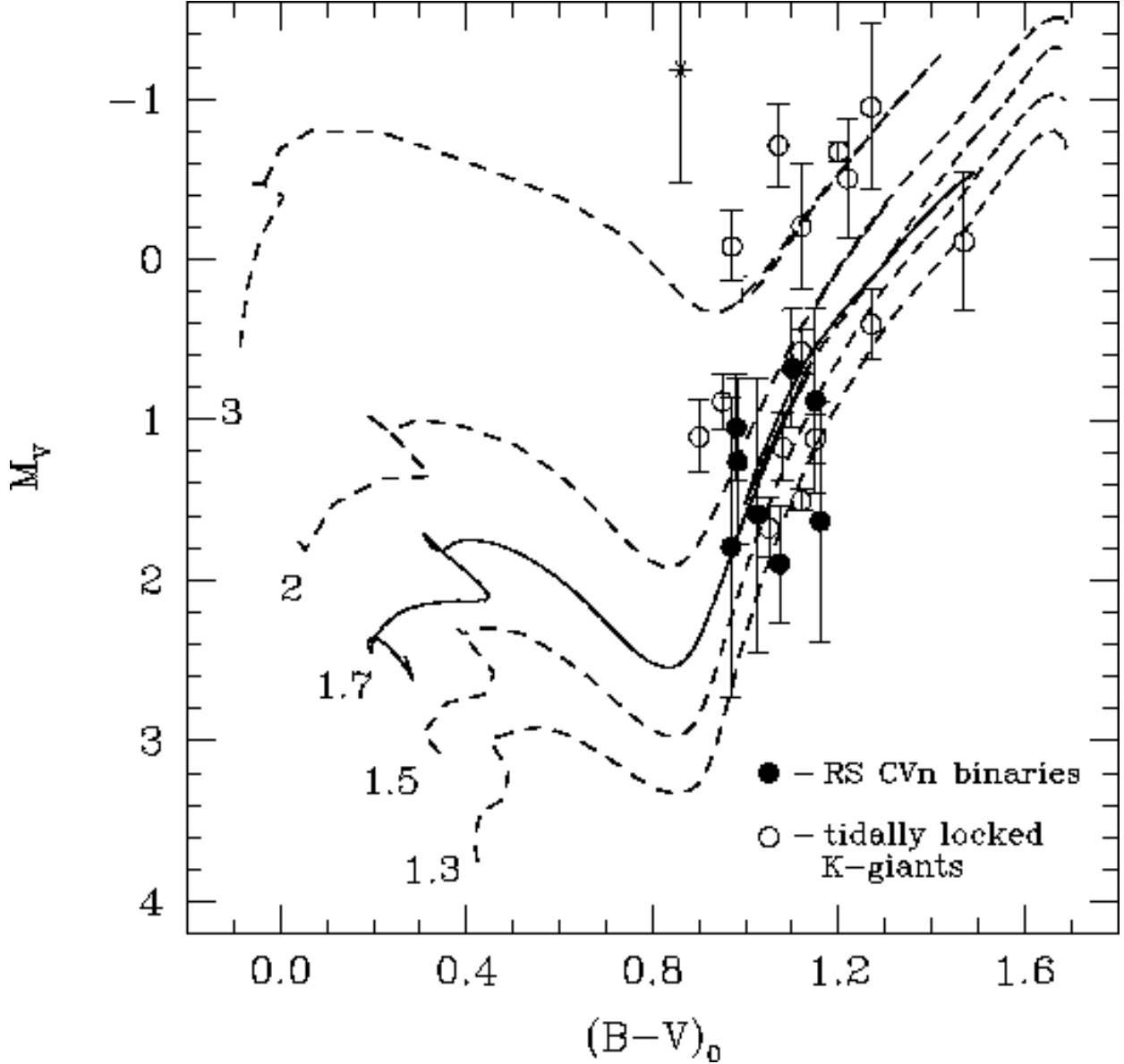}
\caption{{\it Dashed curves} are evolutionary tracks of single solar-metallicity stars.
         Their masses are indicated on the left from the tracks.
         {\it Solid curve} shows the evolution of the 1.7\,$\msol$ star placed into 
         a binary system with $q=0.5$ and $a=80\,R_\odot$. Its initial rotation is
         specified by the parameter $f_\varepsilon = 0.00075$. After having been
         tidally spun up on the lower RGB, this binary star makes an extended bump luminosity zigzag
         between $M_V\approx 0.7$ and $M_V\approx 1.5$. {\it Filled circles} are
         RS CVn binaries that have nearly solar metallicities and
         estimated masses $1.5\la\mstar/\msol\la 1.7$ (\citealt{fea02,mea04,fh05}).
         Each of them has almost equal rotational and orbital periods that correspond
         to $a\approx 50\,R_\odot$ or $a\approx 80\,R_\odot$.
         {\it Open circles} are binary red giants with circularized orbits from Fig.~\ref{fig:f3}.
         {\it Asterisk} is the star HD\,21018 that lies above the dashed curve in Fig.~\ref{fig:f3}.
        }
\label{fig:f9}
\end{figure}

\begin{figure}
\plotone{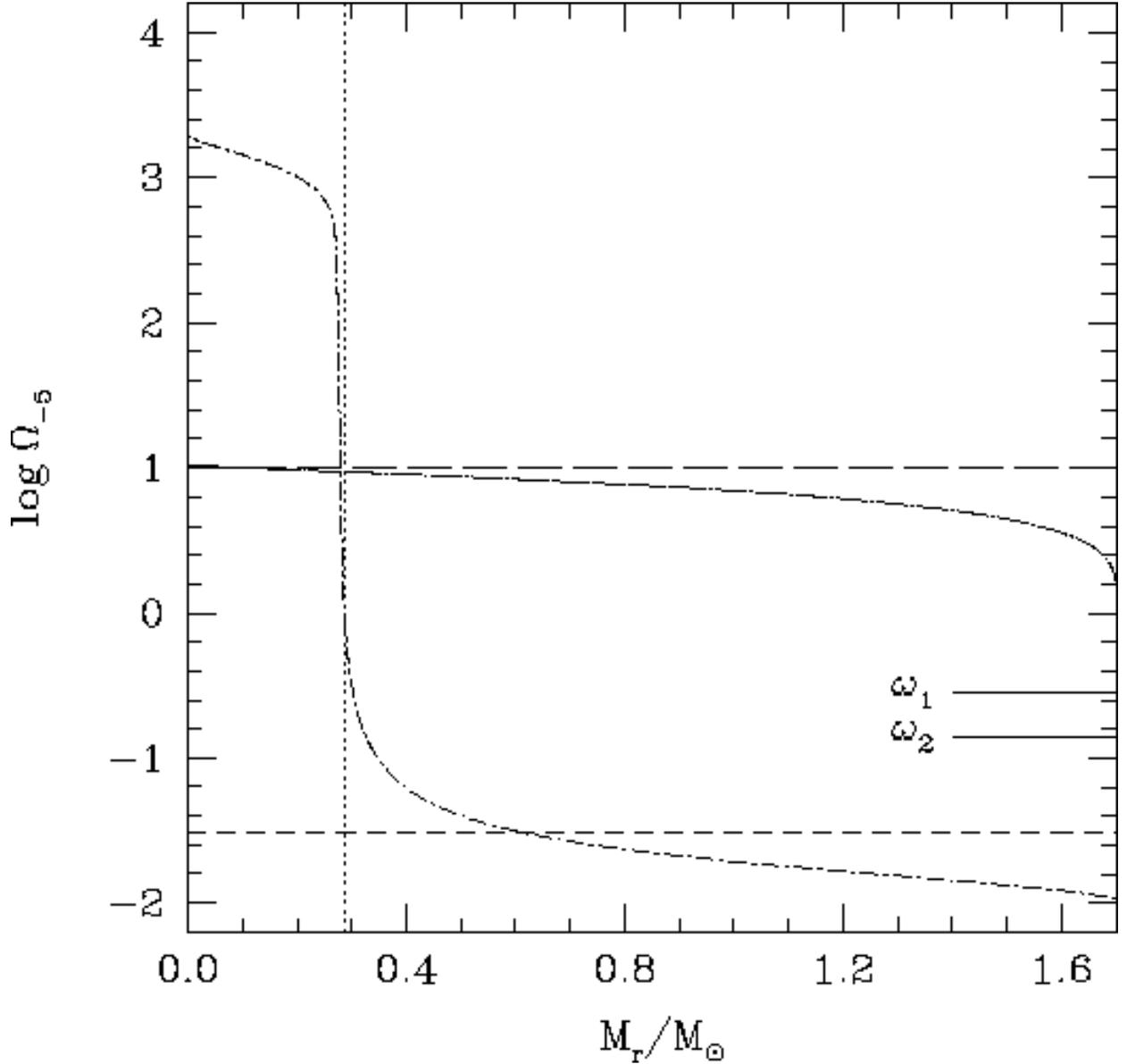}
\caption{{\it Dot-long-dashed} and {\it dot-short-dashed curves} are the ZAMS and
         bump luminosity $\Omega$-profiles in the single model star with 
         $\mstar = 1.7\,\msol$, $Z=0.0188$ and initial rotation parameter $f_\varepsilon = 0.00075$
         rotating differentially (with the specific angular momentum conserved in each mass shell). 
         {\it Long-dashed} and {\it short-dashed curves}
         are the ZAMS and bump luminosity $\Omega$-profiles in the same model but
         for the case of its evolution with uniform rotation starting on the ZAMS with
         $v_{\rm rot}\approx 100$\,\kms. {\it Solid line segments} show the
         orbital angular velocities $\omega_1 = 0.2834$ and $\omega_2 = 0.1400$ (in units of $10^{-5}$\,rad\,s$^{-1}$)
         for binary systems with a 1.7\,$\msol$ primary component, the same mass ratio $q=0.5$ 
         but with different semi-major axes $a=50\,R_\odot$ and $a=80\,R_\odot$, respectively.
         Vertical {\it dotted line} marks location of the BCE in the bump luminosity
         models.
        }
\label{fig:f10}
\end{figure}

\begin{figure}
\plotone{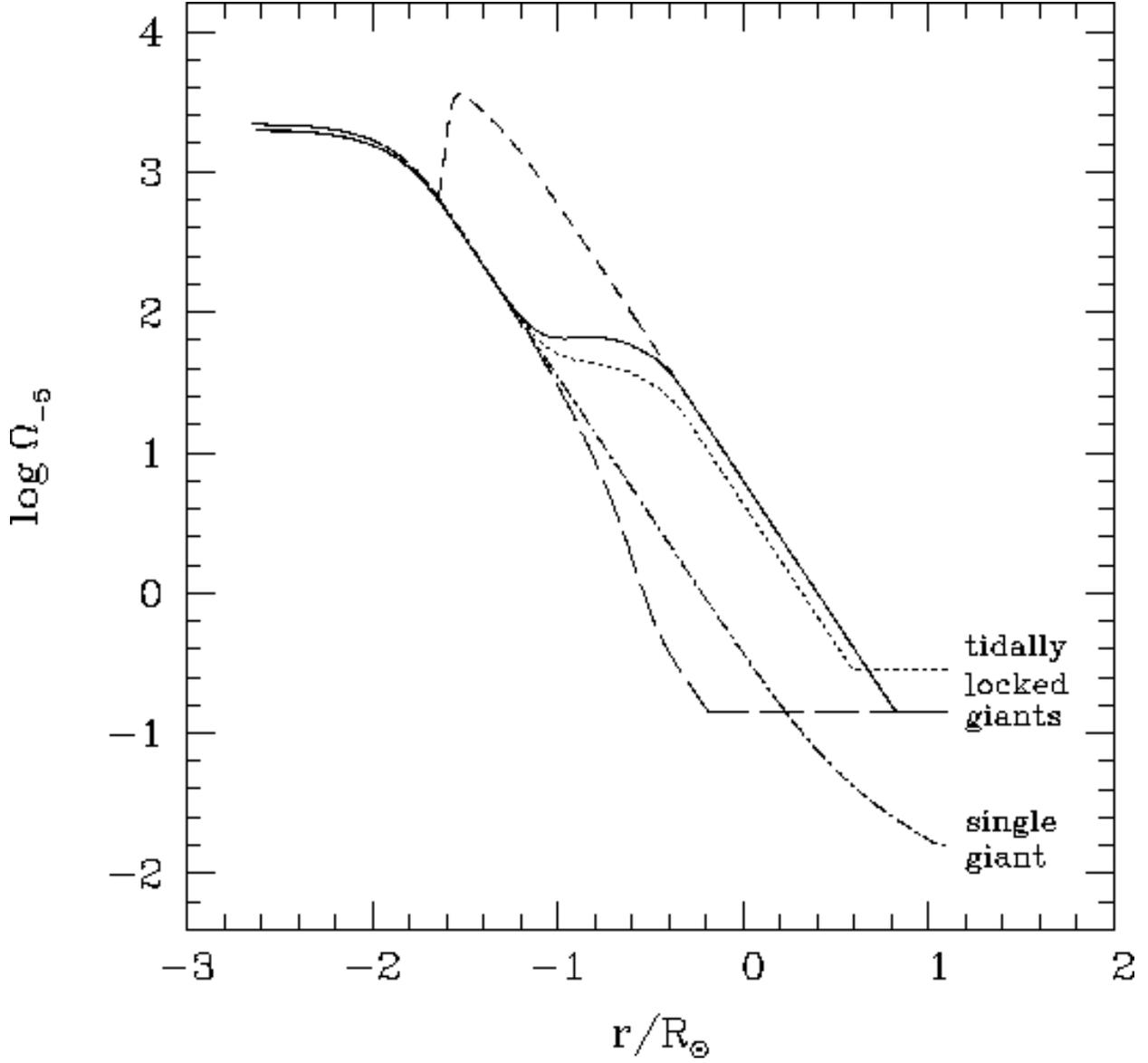}
\caption{Bump luminosity and a later $\Omega$-profiles in differentially
         rotating star with $\mstar = 1.7\,\msol$, $Z=0.0188$ and initial rotation parameter 
         $f_\varepsilon = 0.00075$. {\it Dot-dashed curve} --- in a single star.
         {\it Solid curve} --- in a binary star with $q=0.5$ and $a=80\,R_\odot$.
         It is assumed that only an upper part of convective envelope is
         spun up by the tidal force (for details, see text).
         {\it Short-dashed curve} --- a later profile evolved from the solid one.
         {\it Long-dashed curve} --- in the same binary star but under the assumption
         that rotation of the whole convective envelope has been tidally synchronized.
         {\it Dotted curve} --- a binary star with $q=0.5$ and $a=50\,R_\odot$.
        }
\label{fig:f11}
\end{figure}

\begin{figure}
\plotone{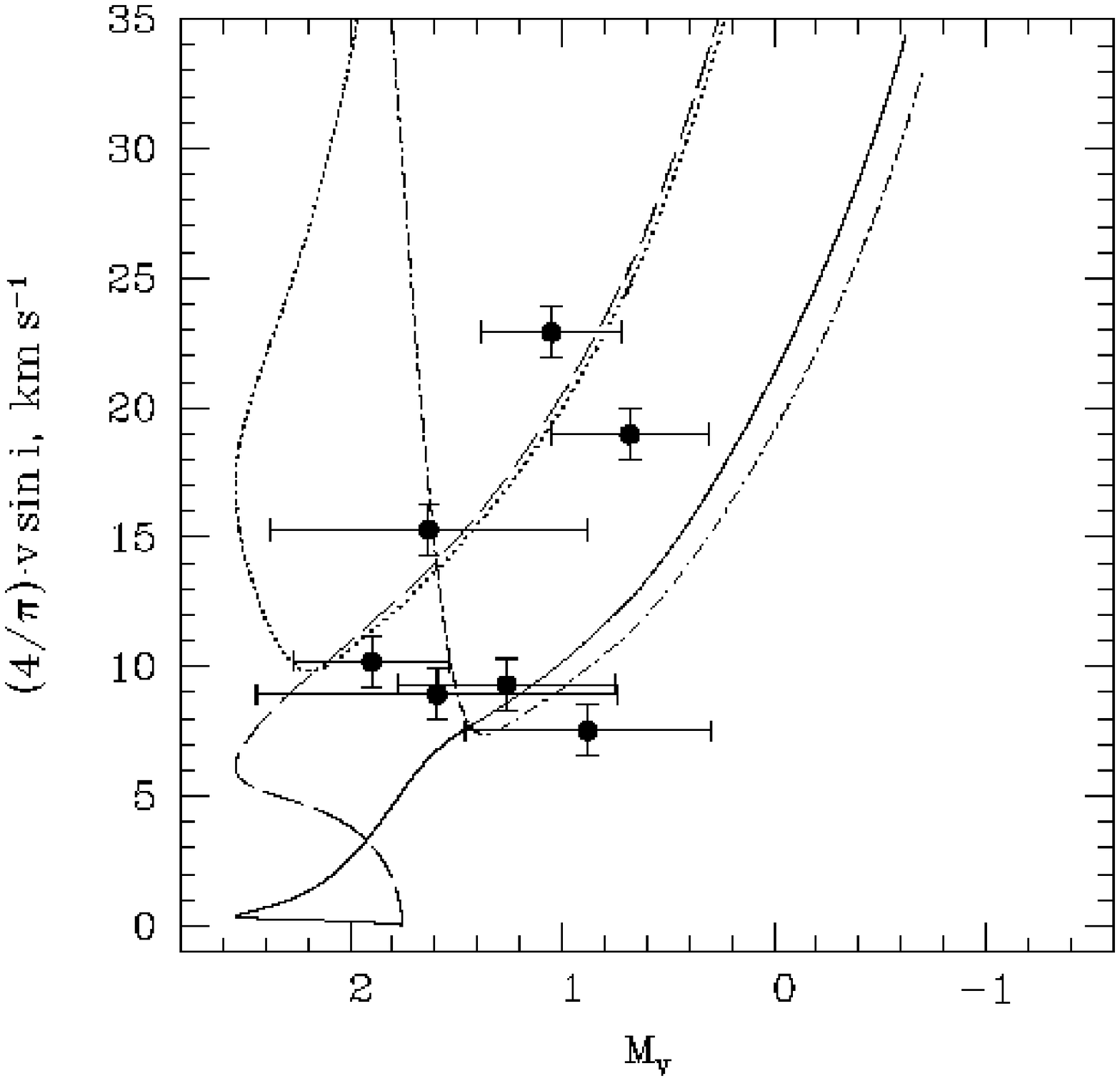}
\caption{Theoretical dependences of the surface rotational velocity $v_{\rm rot}$
         on $M_V$ for the solar-metallicity 1.7\,$\msol$ primary star in binaries with
         the same $q=0.5$ but different semi-major axes: $a = 50\,R_\odot$ 
         ({\it dashed} and {\it dotted curves}) and $a=80\,R_\odot$ ({\it solid} and
         {\it dot-dashed curves}). These
         are compared with locations of RS CVn binaries ({\it circles} are
         data from \citealt{fea02,mea04,fh05}). {\it Solid and dashed curves} start with a subgiant model
         having $R=3\,R_\odot$ and $\Omega = 0$, while
         {\it dot-dashed and dotted curves} initially (at $R=3\,R_\odot$) have $\Omega_{-5} = 2.5$, which
         extrapolates to $v_{\rm rot}\approx 100$\,\kms\ back on the ZAMS.
         The evolution of $v_{\rm rot}=R\,\Omega$ and $a$ due to the tidal interaction is followed by solving equations
         (\ref{eq:domdt}\,--\,\ref{eq:dadt}) as described in text.
        }
\label{fig:f12}
\end{figure}

\begin{figure}
\plotone{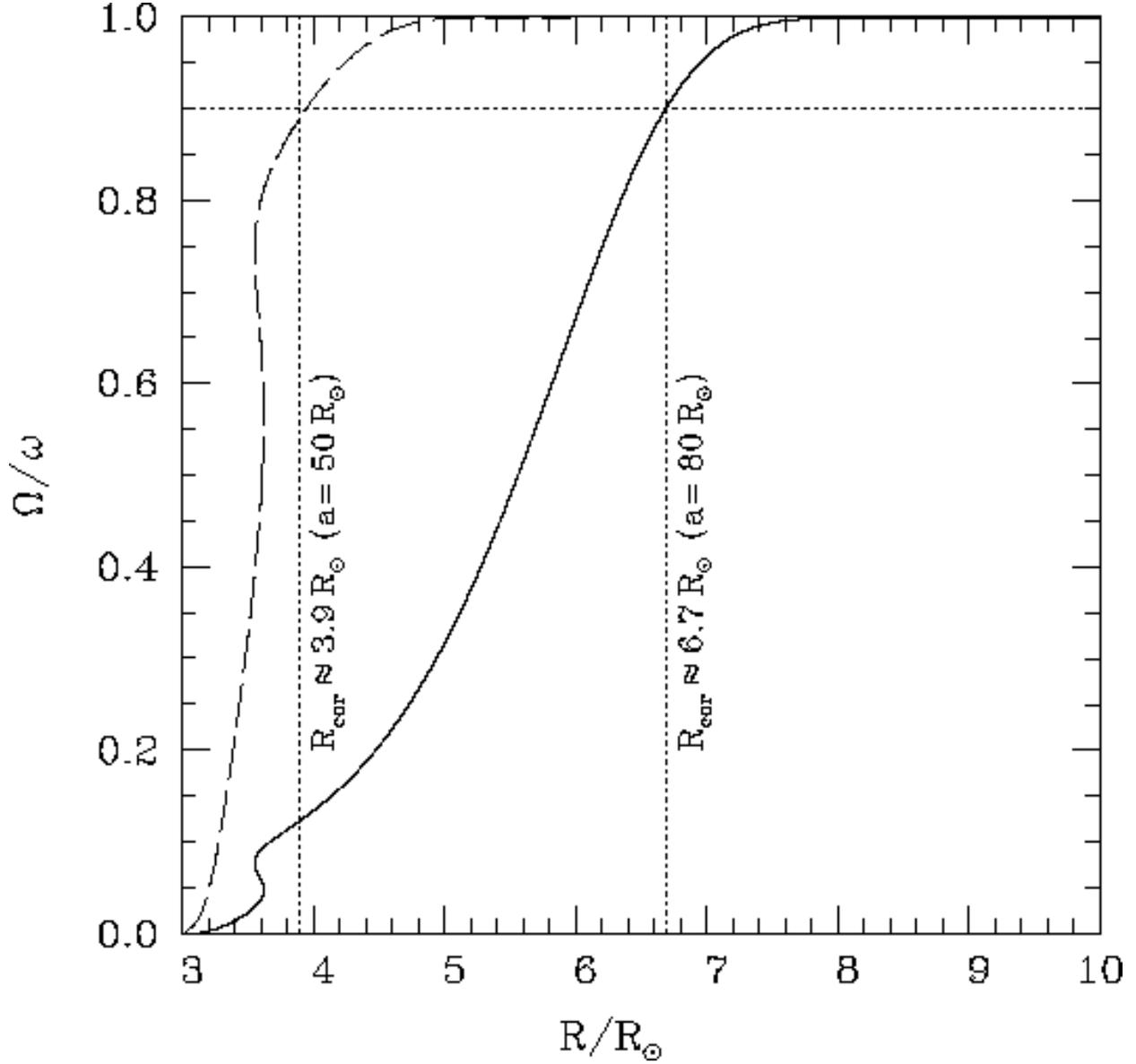}
\caption{Tidal evolution of the ratio of the spin and orbital angular velocities for
         a solar-metallicity $1.7\,\msol$ primary star in a binary system with the same $q=0.5$
         but different semi-major axes: $a = 50\,R_\odot$ ({\it dashed curve}) and $a=80\,R_\odot$ 
         ({\it solid curve}). We define the corotation radius $R_{\rm cor}$
         as the radius at which $\Omega/\omega = 0.9$. During the subsequent evolution
         of the primary star it is assumed that the part of its convective envelope
         above $R_{\rm cor}$ rotates as a solid body with $\Omega=\omega$,
         while the part at $R_{\rm BCE}\leq r\leq R_{\rm cor}$ still rotates differentially.
        }
\label{fig:f13}
\end{figure}

\begin{figure}
\includegraphics[scale=1.0]{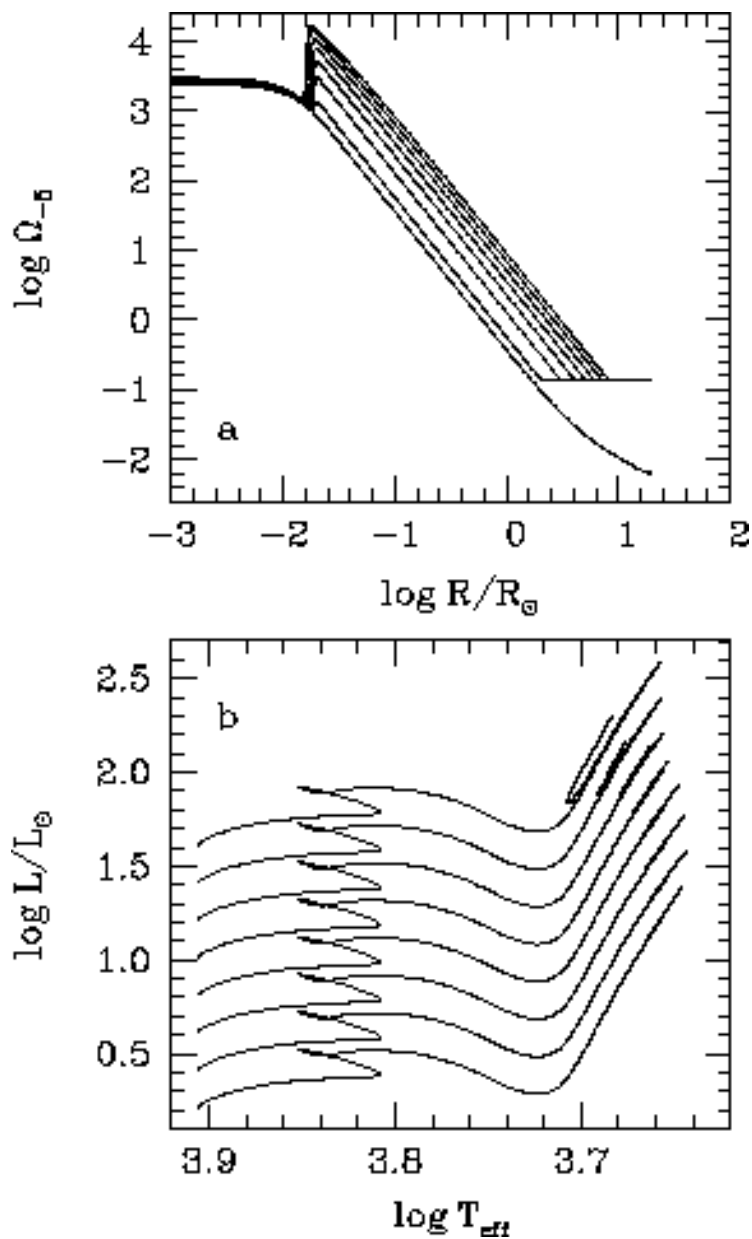}
\caption{The single and binary evolutionary models of a solar-metallicity $1.7\,\msol$ star
         with the corotation radii
         $R_{\rm cor}/R_\odot$\,=\,2, 3, 4, 5, 6, 7, and 8 (from bottom to top).
         The increase of $R_{\rm cor}$ causes the radiative zone to spin up
         (panel a), which makes effects of rotation on the model's internal
         structure to be more pronounced, resulting in a noticeable extension of
         the bump luminosity zigzag (panel b; in order not to overlap each other, 
         the tracks are shifted along the ordinate). 
        }
\label{fig:f14}
\end{figure}

\begin{figure}
\plotone{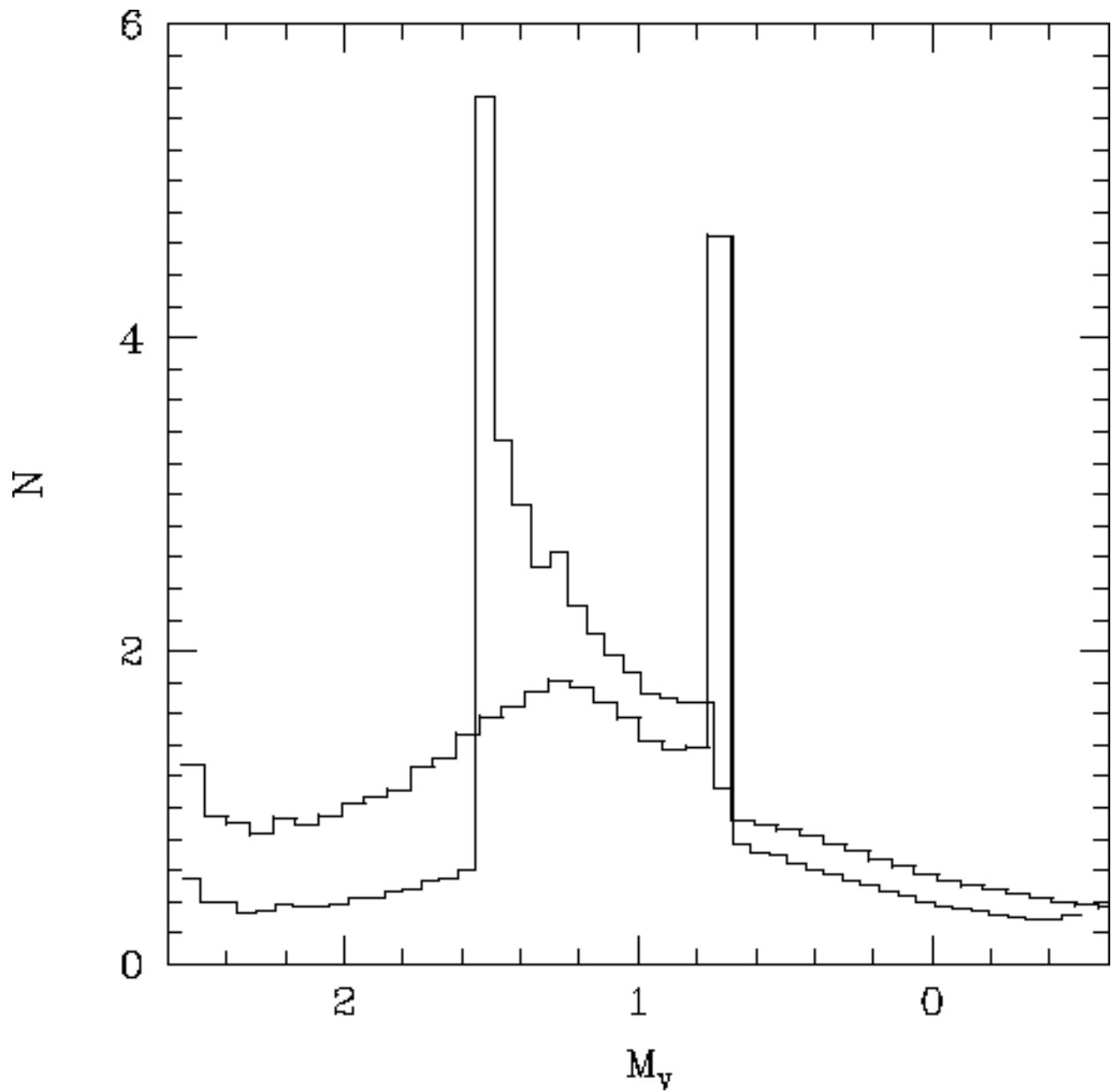}
\caption{Differential luminosity functions constructed using the evolutionary track of
         the single solar-metallicity 1.7\,$\msol$ star ({\it thick curve}) and the track of
         the same star placed into the binary system with $q=0.5$ and $a=80\,R_\odot$ ({\it thin curve}).
         The time spent by the single star between $M_V=2$ and $M_V=0$ is 109 million years, 
         while for the binary star it takes
         146 million years to make an extended zigzag on the CMD when $M_V$ increases from
         0.71 to 1.52 and then decreases back to 0.71. For the binary star,
         the bump luminosity is shifted towards the lower RGB by $\Delta M_V\approx 0.8$.
        }
\label{fig:f15}
\end{figure}

\begin{figure}
\plotone{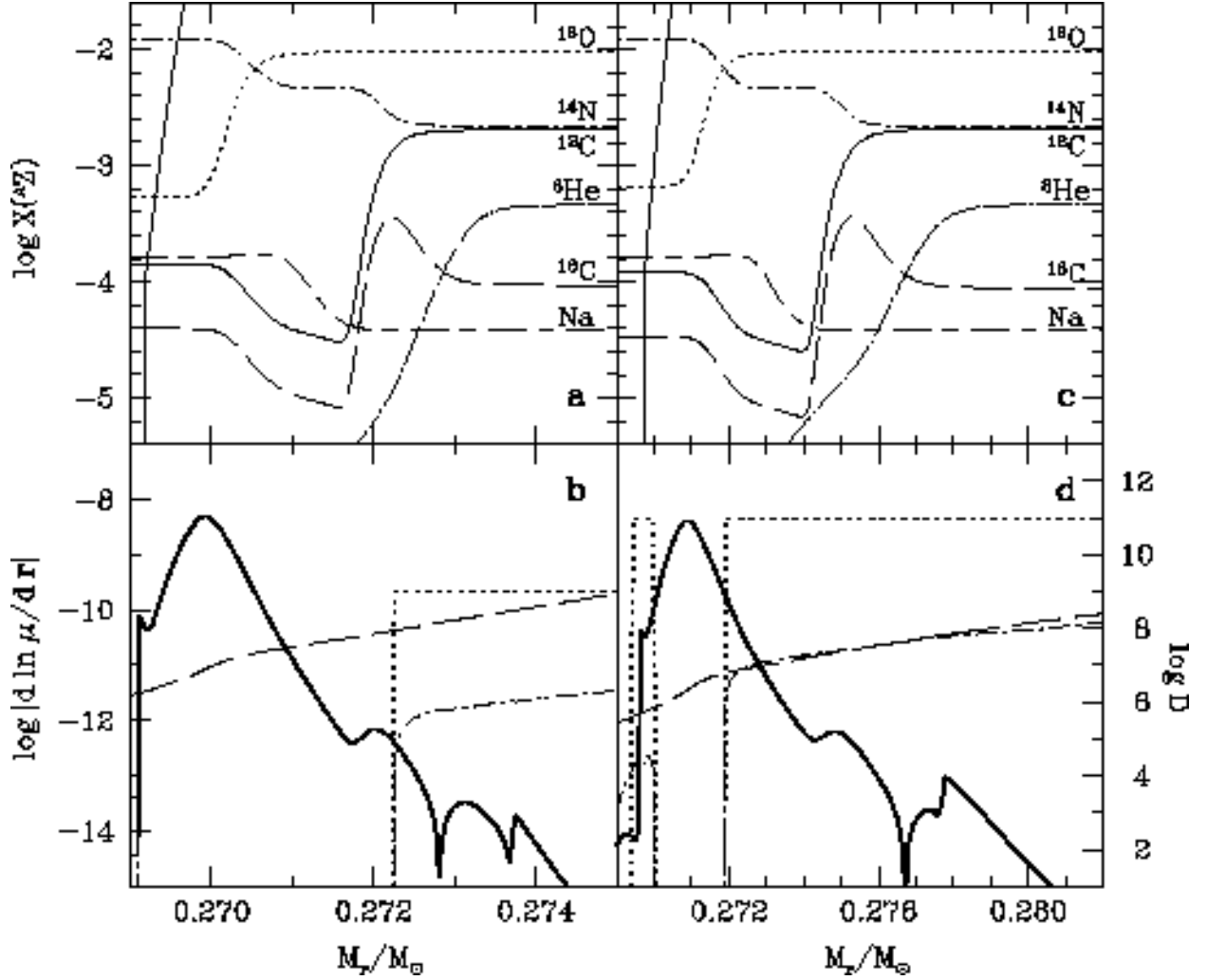}
\caption{Profiles of some element abundances (panels a, c) and diffusivities
         (panels b, d) in the single solar-metallicity 1.7\,$\msol$ model star (panels a, b)
         and in the same model placed into the binary system with $q=0.5$ and $a=80\,R_\odot$
         (panels c, d). Both models are beyond the bump luminosity. In panels b and d,
         {\it dot-dashed curves} show the turbulent diffusion coefficient (\ref{eq:dv})
         for $f_{\rm v}=1$, {\it dashed curves} --- thermal diffusivity,
         {\it solid curves} --- $\mu$-gradients, and {\it dotted curves}
         --- constant diffusion coefficients for canonical ($\dm = 10^9$\,\cs,
         panel b) and enhanced extra mixing ($\dm = 10^{11}$\,\cs, panel d).
         The mixing depth is always constrained by equation (\ref{eq:dvmmix}).
        }
\label{fig:f16}
\end{figure}

\begin{figure}
\plotone{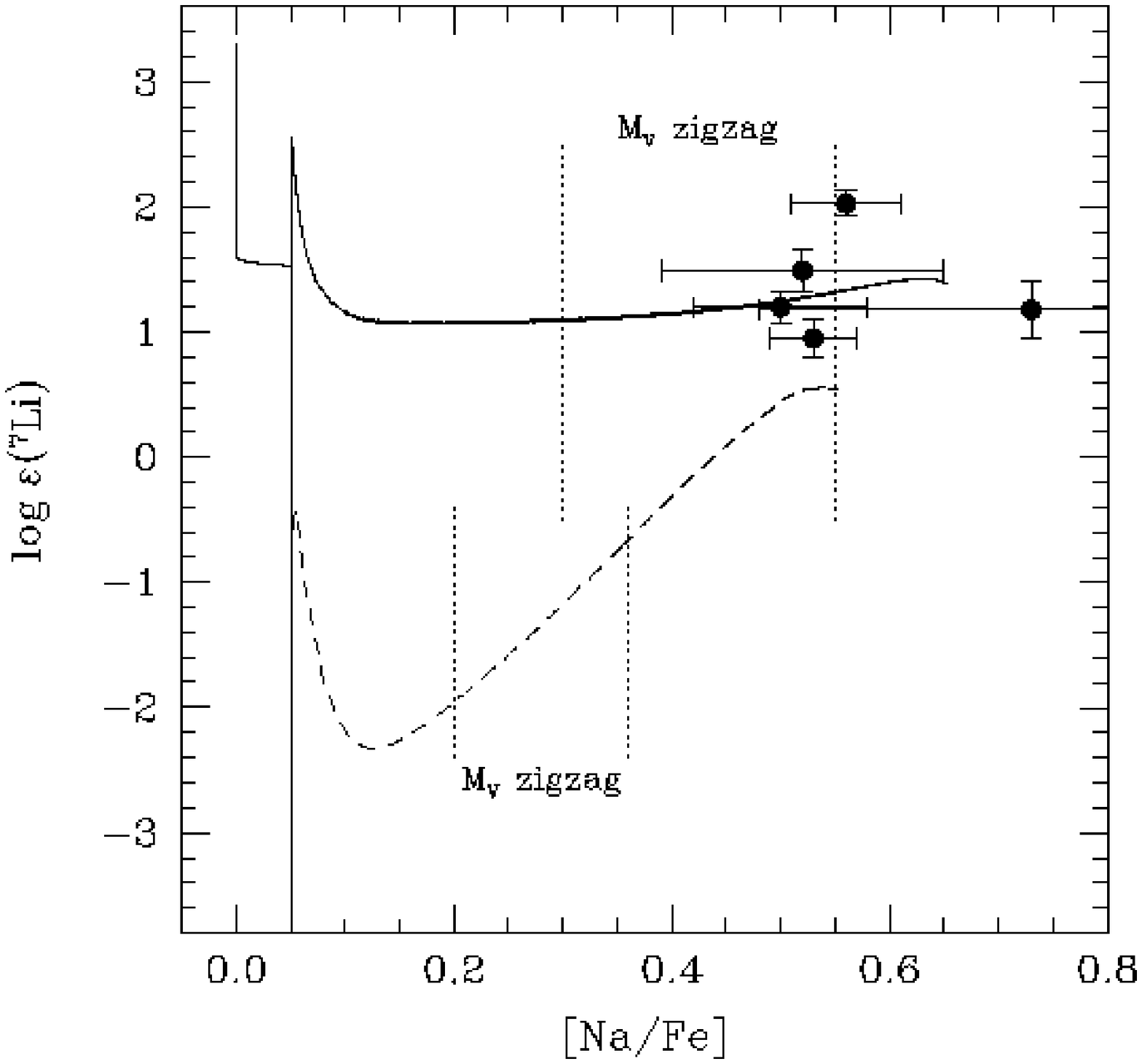}
\caption{Peculiar combinations of the Li and Na abundances in RS CVn binaries
         ({\it circles}), as reported by \cite{mea04},
         and their theoretical reproduction with our model of enhanced extra mixing
         in the solar-metallicity 1.7\,$\msol$ binary red giant. {\it Dashed curve} is obtained
         using the diffusion coefficient (\ref{eq:dv}) with $f_{\rm v}=20$, {\it solid curve} --- for
         the constant coefficient $\dm = 10^{11}$\,\cs\ with the mixing depth from
         equation (\ref{eq:dvmmix}). {\it Dotted line segments} delimit regions of
         the bump luminosity zigzag extended by fast rotation.
        }
\label{fig:f17}
\end{figure}

\begin{figure}
\plotone{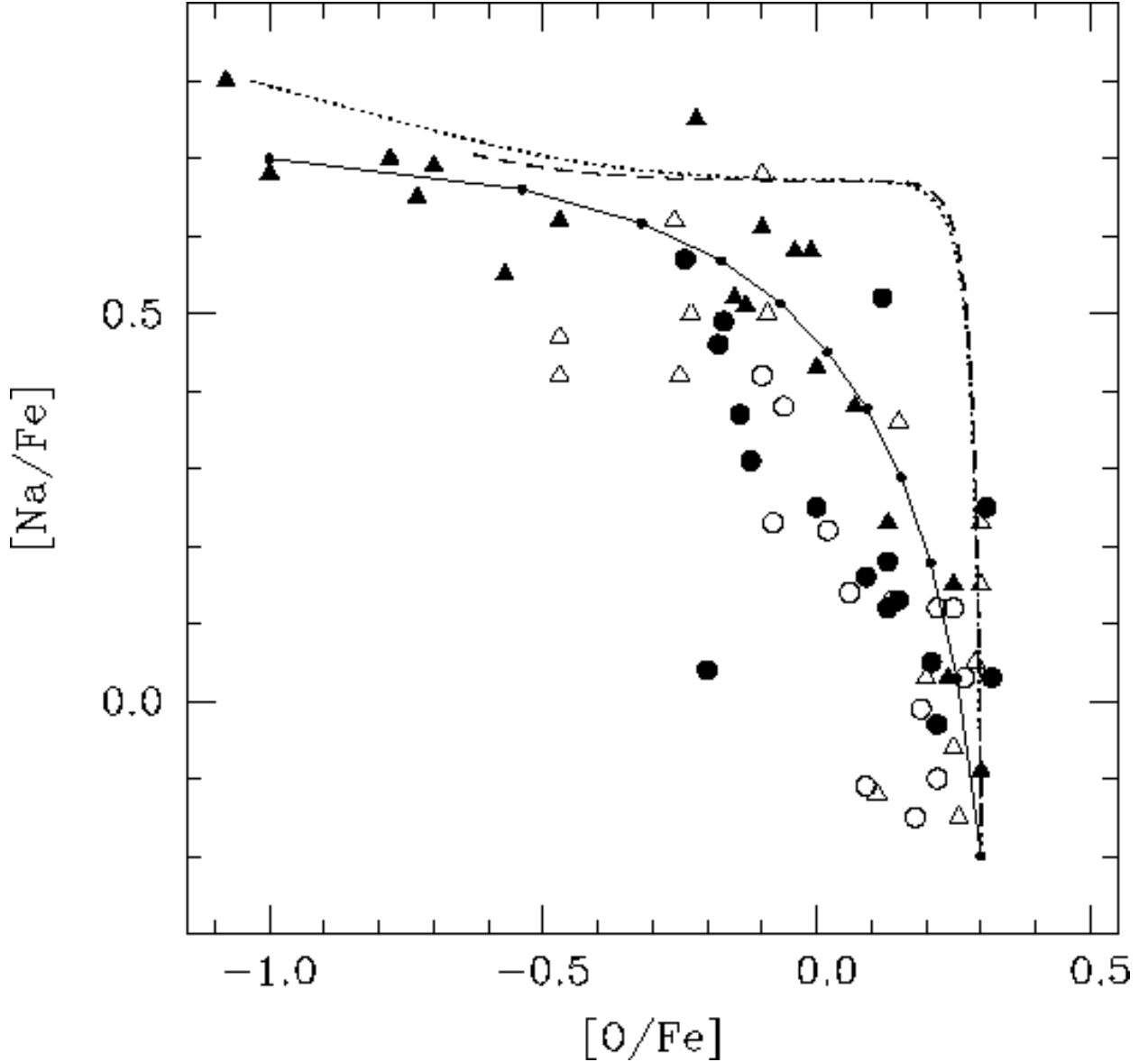}
\caption{The O--Na anticorrelation in red giants in the globular clusters M3 ({\it circles})
         and M13 ({\it triangles}) from a recent update by \cite{sea04}. {\it Filled
         symbols} --- stars with lower surface gravities (higher
         luminosities). {\it Solid curve} shows a simple mixture of a $(1-x)$-fraction
         of material with the abundances [O/Fe]$_{\rm init} = 0.3$,
         [Na/Fe]$_{\rm init} = -0.2$ and a $x$-fraction of material with
         [O/Fe]$_{\rm acc} = -1.0$, [Na/Fe]$_{\rm acc} = 0.7$
         ({\it thick dots} on the curve mark values of $x$ from 0 to 1 with the increment 0.1).
         {\it Dashed curve} is obtained with our model of tidally enforced enhanced extra mixing
         in a red giant with $M=1.0\,\msol$, $Y=0.24$, $Z=0.0005$ and
         $f_\varepsilon = 0.0003$ in a binary system
         with $q=0.3$ and $a=50\,R_\odot$. It is computed using the diffusion
         coefficient (\ref{eq:dv}) with $f_{\rm v} = 20$.
         {\it Dotted curve} presents a test case in which $\Omega$ in the radiative zone 
         has been increased by the factor of 25 as compared to the single star case.
        }
\label{fig:f18}
\end{figure}

\begin{figure}
\plotone{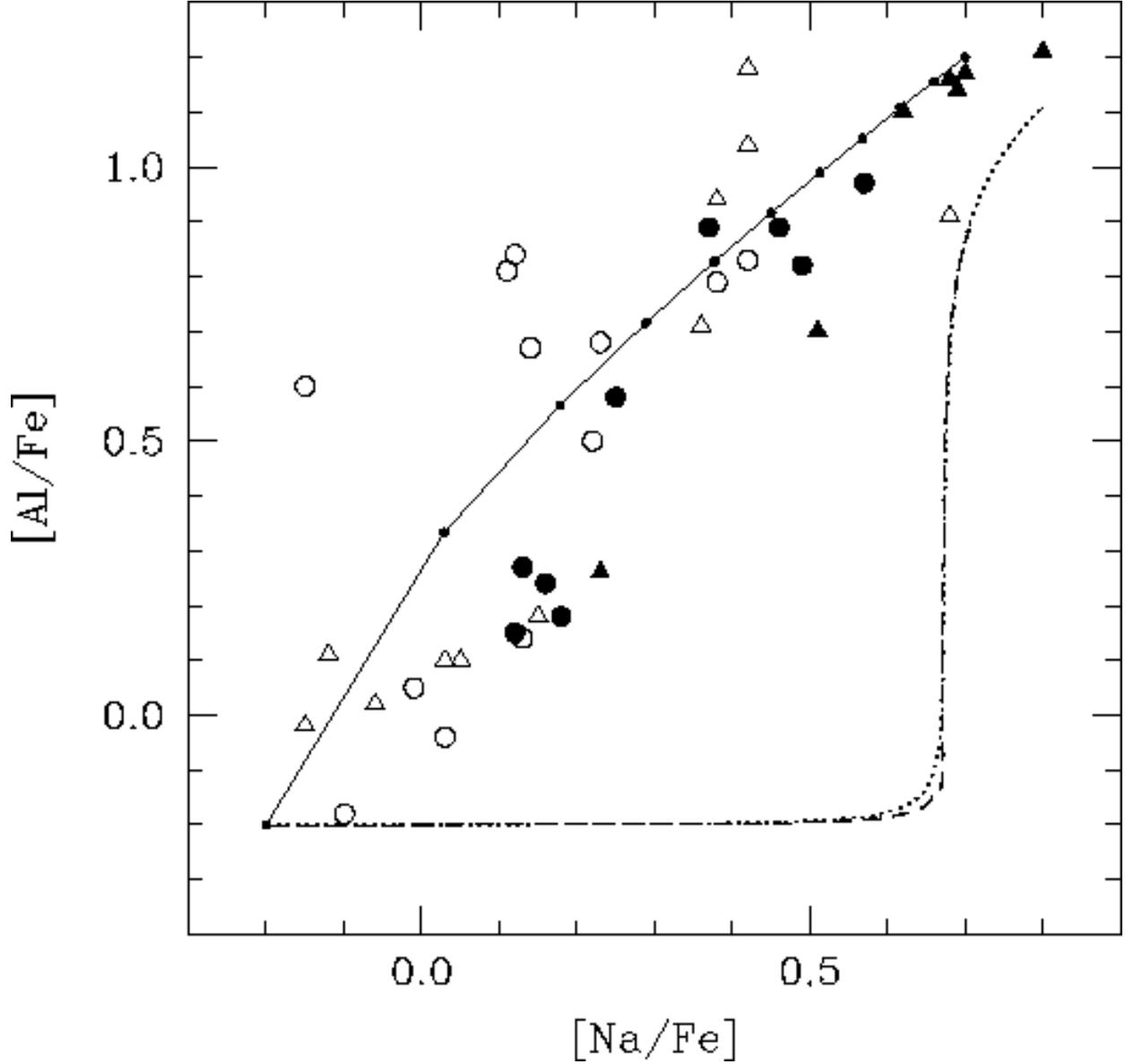}
\caption{The same as in Fig.~\ref{fig:f18} but for the Na--Al correlation.
         For Al, the initial and most extreme abundances are [Al/Fe]$_{\rm init} = -0.2$,
         [Al/Fe]$_{\rm acc} = 1.2$. Following \cite{dt00}, we have additionally assumed that
         [$^{25}$Mg/Fe]$_{\rm init} = 1.2$ and the rate of reaction
         $^{26}$Al$^{\rm g}$(p,$\gamma)^{27}$Si is a factor of $10^3$ as fast as
         its rate from \cite{aea99}.
        }
\label{fig:f19}
\end{figure}

Second, although we do believe that the depth of extra mixing is constrained by equation (\ref{eq:dvmmix}),
when computing its rate, we still cannot choose between the constant diffusion coefficient ($\dm\approx 10^9$\,\cs\ for
canonical and $\dm\approx 10^{11}$\,\cs\ for enhanced extra mixing) and the turbulent diffusion
coefficient (\ref{eq:dvmix}) that is proportional to the radiative diffusivity $K$. The first coefficient
better reproduces the Na--Li data for RS CVn binaries and, unlike the second one, it actually results in a steeper decline of the \cc\
ratio immediately after the bump luminosity as appears to be dictated by observations (e.g., compare the $\log$\,\cc\ panels
in our Fig.~\ref{fig:f2} and in Fig.~4 from \cite{dv03a}). Fortunately, one probably can discriminate between
the diffusion coefficients by measuring the Li abundance in RGB tip and clump stars. As we have predicted,
in the case of $\dm\propto K$ these evolved stars should have values of $\eli\ga 1$ comparable with the post-first-dredge-up values
or even exceeding them. Tentatively, this theoretical prediction seems to have been already confirmed by
\cite{pea01}. However, we think more observational work should be done in this direction.

Finally, we surmise that red giant components of the RS CVn binaries experience enhanced extra mixing caused by
their tidal spin-up. We think that the large Na abundances found in these stars by \cite{mea04} support 
this idea. We did not mention before that the same authors had also reported high Al abundances
in the RS CVn binaries. It is interesting that a similar Na--Al correlation commonly inheres in
the star-to-star abundance variations in globular clusters.
One of its possible interpretations also employs a model of tidally enforced enhanced extra mixing in a binary red giant
(\S\,4.2.2). If it could be possible to corroborate our hypothesis of enhanced extra mixing
in the RS CVn binaries observationally that would also prove that a large amount of Al can be produced
at the same physical conditions under which Na is synthesized (this actually requires higher rates
for the MgAl-cycle reactions responsible for the Al production). We propose to determine
\cc\ ratios and/or CN abundances in the RS CVn binaries to see whether they are consistent with
those resulting from enhanced extra mixing.

\acknowledgements PAD appreciates the generous support of his 1-year
stay in Dartmouth College made available by Prof. Brian Chaboyer
through his research grants. He also thanks the Dartmouth College
staff for their warm hospitality. Research supported in part by a NSF
CAREER grant 0094231 to BC.  BC is a Cottrell Scholar of the Research
Corporation.

                                                                                                                                                          
\clearpage

                                                                                                                                                          
\clearpage
                                                                                                                                                          

                                                                                                                                                          
\clearpage
                                                                                                                                                          
                                                                                                                                                          
\clearpage
                                                                                                                                                          

                                                                                                                                                          
\clearpage
                                                                                                                                                          
                                                                                                                                                          
                                                                                                                                                          
\clearpage
                                                                                                                                                          
                                                                                                                                                          
                                                                                                                                                          
\clearpage
                                                                                                                                                          
                                                                                                                                                          
                                                                                                                                                          
\clearpage
                                                                                                                                                          
                                                                                                                                                          

\clearpage



\clearpage



\clearpage



\clearpage



\clearpage



\clearpage



\clearpage



\clearpage



\clearpage



\clearpage



\clearpage


\end{document}